\documentclass[times,review,preprint,authoryear]{elsarticle}

\usepackage{medima}
\usepackage{framed,multirow}
\usepackage{multicol}
\usepackage{float}
\usepackage{amssymb}
\usepackage{latexsym}
\usepackage{caption}
\usepackage{url}
\usepackage{xcolor}
\usepackage{hyperref}
\usepackage{lineno}
\usepackage{mathtools,xparse}

\journal{}

\begin{document}

\verso{Michail Mamalakis \textit{et~al.}}

\begin{frontmatter}

\title{A novel framework employing deep multi-attention channels network for the autonomous detection of metastasizing cells through fluorescence microscopy}
\tnotetext[tnote1]{An autonomous detection of metastasizing cells }

\author[1,2,3]{Michail \snm{Mamalakis}\corref{cor1}}
\cortext[cor1]{Corresponding authors: 
 Tel.: +44 (0) 7562 576321 ;  
 email: m.mamalakis@sheffield.ac.uk, or mm2703@cam.ac.uk; } 
\author[4]{Sarah C. \snm{ Macfarlane}}
\author [1,5] {Scott V. \snm{ Notley}}
\author [1,4,6,7] {Annica K. B. \snm{Gad}}
\author[1,5]{George \snm{Panoutsos}}

\address[1]{Insigneo Institute for \emph{in-silico} Medicine, University of Sheffield, Sheffield, UK.}
\address[2]{Department of Infection, Immunity and Cardiovascular Disease, and Department of Computer science, Sheffield, UK}
\address[3]{Department of Psychiatry, Cambridge University, Cambridge UK}
\address[4]{Department of Oncology and Metabolism, The Medical School, University of Sheffield, Sheffield, UK.}
\address[5]{Department of Automatic Control and Systems Engineering, University of Sheffield, Sheffield, UK.}
\address[6]{Madeira Chemistry Research Centre, University of Madeira, Funchal, Portugal}
\address[7]{Department of Oncology-Pathology, Karolinska Institutet, Stockholm, Sweden}

\received{June 2023}
\communicated{M. Mamalakis}

\begin{abstract}
We developed a transparent computational large-scale imaging-based framework that can distinguish between normal and metastasizing human cells. The method relies on fluorescence microscopy images showing the spatial organization of actin and vimentin filaments in normal and metastasizing single cells, using a combination of multi-attention channels network and global explainable techniques. We test a classification between normal cells (Bj primary fibroblast), and their isogenically matched, transformed and invasive counterpart (BjTertSV40TRasV12). Manual annotation is not trivial to automate due to the intricacy of the biologically relevant features.  In this research, we utilized established deep learning networks and our new multi-attention channel architecture. To increase the interpretability of the network - crucial for this application area - we developed an interpretable global explainable approach correlating the weighted geometric mean of the total cell images and their local GradCam scores. The significant results from our analysis unprecedently allowed a more detailed, and biologically relevant understanding of the cytoskeletal changes that accompany oncogenic transformation of normal to invasive and metastasizing cells. We also paved the way for a possible spatial micrometre-level biomarker for future development of diagnostic tools against metastasis (spatial distribution of vimentin).}
\end{abstract}


\end{frontmatter}

\section{Introduction}
Metastasis is the main cause of the death of cancer patients \citep{1}. The change from a healthy and stationary to a motile and metastasizing cell is accompanied by an altered spatial organisation of the cytoskeleton cells.
The cytoskeleton is a filamentous network of protein polymers, which  which governs the mechanical, adhesive and motile properties of cells. The actin microfilament system and the intermediate filament protein vimentin are key components of this cytoskeleton which, when  defective, promote deregulated motility of metastasizing cells \citep{3,4}. Vimentin has been used in the clinic for more than half a century to diagnose invasive carcinoma, and distinguish it from stationary carcinoma \textit{in situ}, and regulators of actin filaments are often cancer-causing oncogenes \citep{5}. We have previously observed that the spatial distribution of the vimentin and actin filaments are altered in metastasizing cells, as compared to normal control cells \citep{m1a,m2a}. Given the key role of vimentin and actin filaments in metastasis, more detailed and objective understanding of the spatial organisation of the filaments is altered in metastasizing cells can inform future development of anti-cancer diagnostic and therapeutic tools \citep{7,8}.

Artificial intelligence and metastatic cell classification has gained interest in the last decade with a number of scientific conferences hosting related studies covering associated relevant subjects and topics \citep{ieee1,ieee5,mic1,mic2}. Because the characteristic patterns found in cell scanning are not uniform, there is a high likelihood of human failure in the classification of metastatic cells, which can potentially lead to significant mistakes in the detection and treatment of cancer \citep{ieee1,ieee2}. In light of this, artificial intelligence approaches have shown promise in assisting experts in the diagnosis of cancer metastasis \citep{ieee4,ieee5}. However, in the majority of metastatic cell classification work there is a lack of accurate explanation in the classification process and/or the detection rules that the methods infer from the data. The lack of explainability of the decision process impedes the usefulness of such methods, with respect to knowledge discovery, and has ethical and legal implications in the uptake of such methods into mainstream clinical practice \citep{mx3}.  

The aim of this study is to establish an interpretable artificial intelligence image-based methodology that can distinguish between normal and metastasizing cells based on the organisation of filamentous actin and vimentin. Overall system interpretability is pivotal here, to establish the underpinning biological knowledge as well as confirm the biology-relevant forecasting capability.  To deliver these objectives, we used a variety of different established networks, as well as our new deep attention network and a combination of local explainable techniques (GradCam) with an interpretable generalized technique (weighted geometric mean) to study the pattern learning of the networks. Five different established deep learning network architectures (Densenet-121, Resnet-50, Vgg-16, DenRes-131 and ViT) are assessed in terms of best performance in a binary classification task of normal or metastasizing cells. We compare the performance of the established networks, with our new developed multi-attention channels networks, to take into consideration all permutations of the colour channels (RGB). This level was compulsory as, in contrast to the analysis of standard colour images with a high level of spatial correlation across channels, in this case each channel relates to different information of the cell's physiology (DNA, RNA, shape etc). The approach adopted in this work is multi-spectral, in the sense that three separate images, pertaining to the microscopic structure of actin, vimentin and DNA, are collected for each cell. To acquire these images, we have used high-content fluorescence microscopy. All imaging parameters, including focus strategy and LED intensity, can be stored which allows for highly reproducible multichannel imaging. For each cell, the three image sources are combined into the RGB channels of a single image to generate a cohort of 100 normal cell and 120 oncogenically transformed, invasive, metastasizing cells.
\noindent Our main contributions are:
\begin{itemize}
\item {A rigorous study of different established and newly developed deep learning models to justify an unbiased highly accurate binary classification.}
\item {The development of new deep learning architectures, by using multi-attention channels mechanisms to force the networks focus on the global spatial correlation across the image channels which relate to actin, vimentin and DNA.}
\item {A statistically principled approach and explainable methods to study a variety of features and decision patterns of the networks.}
\item {The development of a generalised explainable method (weighted geometric mean, Gmean-GradCam) to study the decision pattern of the networks on the classification task.}
\item {Detailed understanding of how the spatial distribution of the cytoskeleton is, when defective, and how it results in metastasizing cells.}
\end{itemize}
\begin{figure*}[h]
\centering 
\centering
\includegraphics[trim={1.15cm 17.cm 5.7cm 3.5cm},clip,scale=1.4]{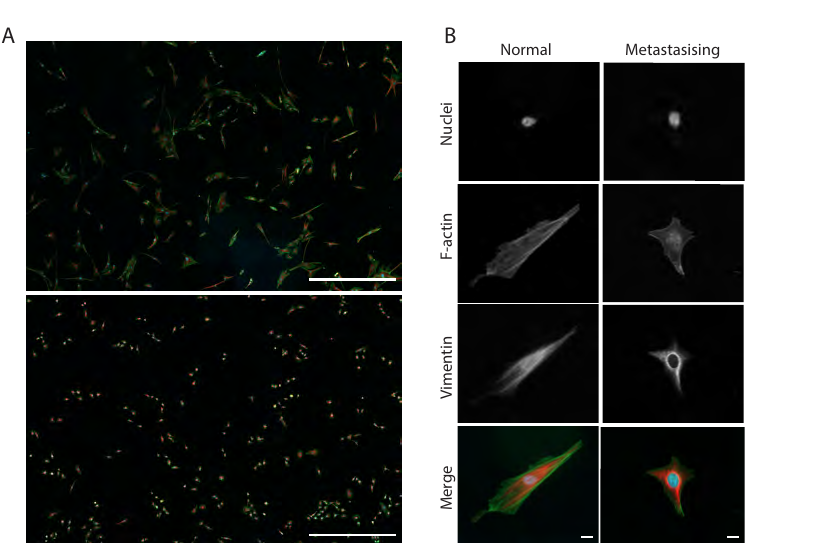}
\caption{Representative images of normal and metastasizing cells. (A) Normal (top) and metastasizing (bottom) cells. Scale bar: 1 mm. (B) Single cells, showing nuclei, F-actin, vimentin, and merged images of nuclei (blue), F-actin (green) and vimentin (red). Scale bar: 20 $\mu$ m. Images show representative cells from at least 2 experimental repeats}
\label{intro}
\end{figure*}

The manuscript follows a structured organization. Firstly, in Section 2, a comprehensive literature review is presented, covering essential topics such as cells classification, explanability, and multi-attention techniques. Subsequently, in Section 3, the methodology employed for the deep learning networks and the corresponding explanability framework are expounded. The description of the experimental setup for extracting cell images, as well as the pertinent training information of the deep learning networks, is presented in Section 4. The results of the experiments are reported in Section 5, followed by a detailed discussion in Section 6.

\section{Related Work}

\subsection{Cells classification}

The classification of metastatic or cancer cells using deep learning techniques has been trending during the last decade \citep{m1,m2,c6,at2}. For example, generative neural networks in combination with supervised machine learning has been used to classify melanoma xenografts as metastatic and non-metastatic cells \citep{c1}. These predictions were further validated in melanoma cell lines with unknown metastatic cells of mouse xenografts and used to generate in silico cell images that amplify the critical predictive cell properties \citep{c1}. Similarly, \citet{c2} developed a DL methodology to help in the diagnosis of leptomeningeal metastasis. This method delivers quick evaluation and accurate detection of cancer cells and cancer's primary source of leptomeningeal metastasis, and can potentially be used to assist in early diagnosis and treatment. The work of \citet{c4} uses a machine learning-based approach for iPS progenitor cell identification. They utilise a prediction model using XGBoost of six types of features and best time windows. Their cross-validation, holdout validation and independent test experiments, show that the morphology and motion pattern of iPS progenitor cells is different from the that of normal cells.  \citet{c7} presented an unbiased phenotypic profiling platform that combines automated cell culture, high-content imaging, cell painting, and deep learning. They apply this platform to primary fibroblasts from 91 Parkinson’s disease patients and matched healthy controls, creating the largest publicly available Cell Painting image dataset to date at 48 terabytes.
\citet{c5} developed a wide and deep learning approach to construct a cell-classification prediction model that can learn specific patterns and ensure the inclusion of biologically relevant features in the final prediction model, and provides higher accuracy when compared to state-of-the-art cell classification algorithms. \citet{c8} trained a deep learning classifier to predict cancer type based on patterns of somatic passenger mutations detected in whole-genome sequencing of 2606 tumours representing 24 common cancer types. They observed that adding information on driver mutations reduced the accuracy and underscored how patterns of somatic passenger mutations encode the state of the cell of origin. Lastly, \citet{c9} presented a convolutional neural pipeline allowing the discrimination between normal and human breast cancer cell,both in single cell culture and in tissue sections. They analysed the sensitivity of the pipeline by detecting subtle alterations in normal cells when subjected to small mechano-chemical perturbations that mimic tumor microenvironments. Their method did not involve any explainable artificial intelligence (XAI) technique, but it provides interpretable features that could aid pathological inspections.

Many studies highlight the importance of interpretability, explainability, multi-modal centre validation, and the quality of imaging scanning systems (X-ray, etc) \citep{c3,c6,mx3}, underlining that only methods that are highly explainable can be approved by the regulatory agencies as AI tools to diagnose cancer. However, in the majority of studies the detection of metastatic cell lack of an accurate explanation of the classification process, and the pattern/rules that the network follows for each decision \citep{ieee1,m3,m4,c2,c8}. In contrast, in this current study, we deliver an analytical explanation of a number of deep learning pipelines and explainable techniques. 

\subsection{Interpretable and explainable techniques}
There has recently been a major increase in the number of publications on XAI and machine learning (XML) \citep{xai,mx1,mx2}. XAI is categorized into two general method techniques; posthoc and transparent. The transparent methods focus on models that have simulatability, decomposability and transparency, and are linked mainly with linear techniques like Baysian classifiers, support vector machines, decision trees, and K nearest neighbour \citep{xai22}. In contrast, posthoc methods are often related to AI techniques that attempt to find nonlinear mappings with high complexity datasets. A common technique is local interpretable model-agnostic explanations (LIME), which tests the robustness of the explanation by adding noise in the dataset perturbating the data \citep{xaimi}. Post-hoc includes techniques focusing on the non-linear behaviour of the model and of the dataset. Thus, post-hoc is a super-group of the model-specific techniques and model agnostic techniques \citep{xai,xai22}. In computer vision, the model agnostic techniques, like local interpretable model-agnostic explanations, perturbation, and layer-wise relevance propagation (LRP) are the most applicable \citep{lime}. The model-specific techniques include methods such as feature relevance \citep{xaims}, condition-based explanation, and rule-based learning \citep{xai22}. The explainable methods for medical imaging focus on techniques of attribution and perturbation \citep{xai_surv}, where the attribution techniques assign any feature from the network a weight definition of the feature output for each layer (LIME, LRP, GRAD-CAM) \citep{xai_surv}, while the perturbation techniques mainly focus on input perturbations with respect of the prediction of the ML or DL techniques \citep{xaimi}. Another important techniques that explores the latent features significant to the model is Occlusion \citep{xai_surv,xaimi}. The most used explainable method In medical imaging is the GRAD-CAM method \citep{xaimi}, which is why we, in this study, examine our networks based on GRAD-CAM.

\subsection{Multi-layer attention networks in deep learning}
Many studies have combined existing deep learning architectures with multi-layer attention models exploiting attention mechanisms to focus on a global space of features \citep{ma5,at1,m5}. For example, \citet{ma1} used a multi-attention strategy that utilises bidirectional long short-term memory networks, with a final step that achieves the classification of RS image scenes with multilabels. The main drawback of this method is the lack of strategy for an adaptive definition of local areas based on the semantic content of RS images and  a data summarization strategy  instead of stacking the local descriptors of the proposed approach. \citet{ma2} utilised a deep learning network with an attention-reference mechanism to solve the few-shot scene classification task in remote sensing. They used both global and local attention mechanisms to improve the performance of the model. However, the sensitivity of the method to different backbone architectures is not studied. Moreover, \citet{ma3} developed a deep learning model with spatial and channel attention modules (CBAM) to capture subtle but distinguishing features. They use features to isolate the background influence in the classification task.

Taken together, the main limitation of the majority of studies is the limited variation in different backbone architectures (thus in established convolutional networks) and in combination with multi-attention layers. In this study we cover a number of different combinations of these two components.

\section{Methods}

\subsection{Network architectures background}
We have used four established networks (VGG-16, DenseNet-121,ResNet-50, DenResCov-19 and ViT) and a new deep learning model (DenRes-131).

\par VGG-16 is an established convolutional neural networks (CNNs) with a combination of pooling and convolution layers \citep{VGG2}. ResNet-$L$ is inspired by the structure of VGG nets \citep{VGG2}. The network comprises of $L$ layers, each of which implements a non-linear transformation. In the majority of ResNet-based networks, the convolutional layers have $3 \times 3$ filters. Downsampling is performed by convolutional layers with a stride of $2$. The last two layers of the network are an average pooling layer, followed by a $1000$-way fully connected (FC) layer. 
The networks uses the feed-forward inputs \(x_{i}\) and the input \(x_{i-1}\) of the previous layer. Thus ResNet adds a skip-connection and the identity function is given by:
\begin{equation}
x_i = H_i(x_{i-1}) + x_{i-1}
\label{1}
\end{equation}
where $H_1$ the layers transformation function.
\par Densenet-$L$ is a convolutional network. The network comprises of $L$ layers, each of which implements a non-linear transformation. These transformations can be different function operations, such as Batch Normalization, rectified linear units (ReLU), Pooling, and Convolution \citep{De}. DenseNet architectures utilise the feature-maps of all previous layers as an input in the $i$th layer. The input of $i$th layer is given by the equation:
\begin{equation}
x_i = H_i\left([x_0, x_1,\cdots, x_{i-1}]\right)
\label{2}
\end{equation}
where $[x_0,x_1,\cdots, x_{i-1}]$ refers to the concatenation of the feature-maps produced in layers $0,\ldots,i-1$.  All inputs of a composite function $H_i(\cdot)$ are concatenated into a single tensor. Each composite function is a combination of batch normalization (BN), followed by a rectified linear unit (ReLU) and a $3 \times 3$ convolution (Conv). 

\begin{figure}
\centering 
\centering
\includegraphics[trim={5.5cm 2.5cm 2.5cm 2.5cm},clip,scale=.385]{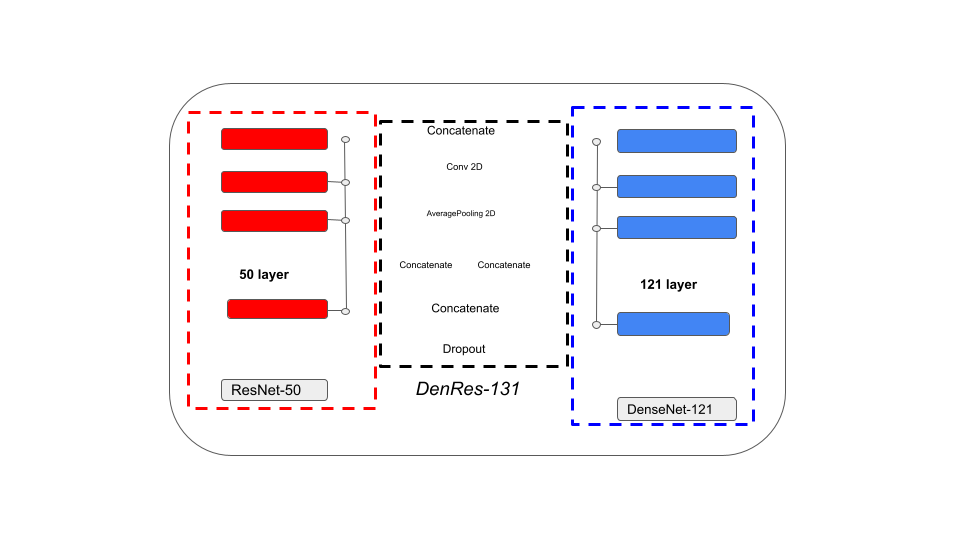}
\caption{The DenRes-131 network a modified version of the DenResCov-19 state of the art network. The DL family (Vgg-16, DenseNet-121, ResNet-50, DenRes-131). }
\label{ma}
\end{figure}

DenRes-131 is a modified version of \citep{our} network which has two dropout layers to estimate the epistemic uncertainty of the model and introduces regularisation to reduce overfitting (Fig. \ref{ma}). We used a probability of 0.3 in both layers. The DenRes-131 combines four blocks from ResNet-50 and DenseNet-121 with width, height, and frames of $58 \times 58 \times 256$, $28 \times 28 \times 512$, $14 \times 14 \times 1024$, and $7 \times 7 \times 2048$, respectively. Each of the four outputs feeds a block of convolution and average pooling layers. The final layer uses a soft-max regression as the final network output resulting in a classification decision.

Finally, we used the state of the art vision transformer (ViT). ViT uses embedded patches to split the initial image into fixed-size patches based on a linear projection of the flatten patches. The main transformer encoder include two levels of layernorm, a multi-head attention layer and a multi-layer perceptor layer \citep{vit}.

\subsection{Multi-head attention encoder using backbone deep learning networks}
We used a combination of multi-head attention layers to focus in the global diversity and variation of the three different channels of the fluorescence microscopy images (R: red, G: green, B: blue). In addition, we checked the performance of different combinations of backbone deep learning architectures combined with different multi-head attention structures. As backbone networks, we chose a three layer CNN network, and ResNet-50 and DenseNet-121 networks (second best, and worst performance CNN networks of this study, respectively), to avoid any bias effect of the chosen backbone networks. 

\begin{figure*}[h!]
\centering 
\centering
\includegraphics[trim={0.cm 0.cm 0.0cm 0.0cm},clip,scale=.7]{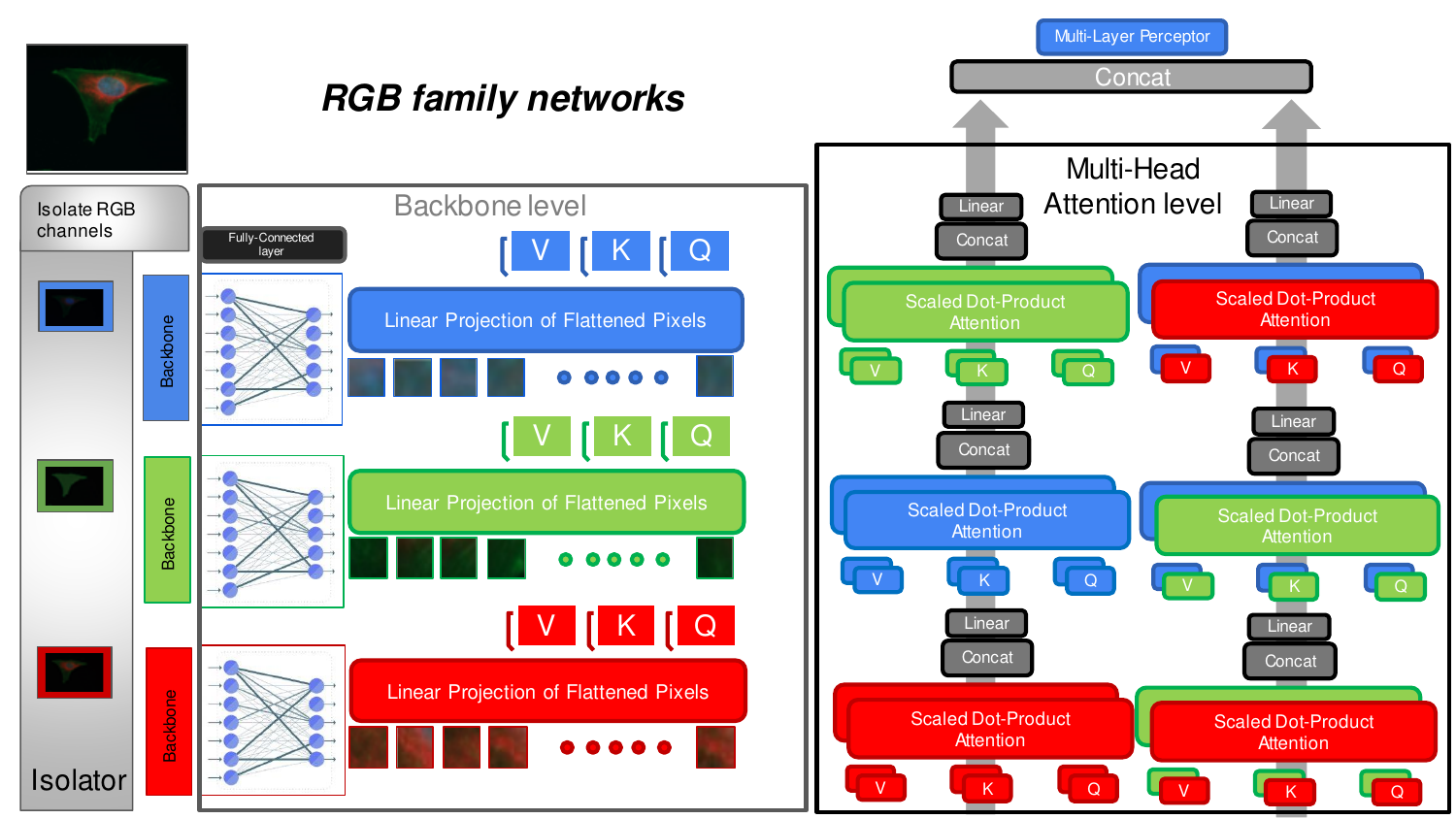}
\caption{The networks and the multi-head layer. The RGB family (RGB, RGB-Res, RGB-Den). The RGB architecture has three main levels isolator, backbone, multi-head attention. The Isolator is the first level which isolate the three channels of input image (Red, Green, Blue). Following the Backbone level uses established classifiers (convolution networks or ResNet-50 or DenseNet-121) which utilise as input each of the three isolated channels. Lastly, the multi-head attention level uses each output from the backbone network (V,K,Q) to initialise a scaled dot-product attention layer. Six multi-head attention layers are used; three self-attention of each channel (Red, Green, Blue) and three pair channels (Red/Green, Blue/Green, Red/Blue) multi-attention layers. following by a concatenation of them and a multi-layer perceptor to classify the input image.}
\label{ma1}
\end{figure*}

\begin{figure*}
\centering 
\centering
\includegraphics[trim={0.0cm 0.cm 0.0cm 0.0cm},clip,scale=.6]{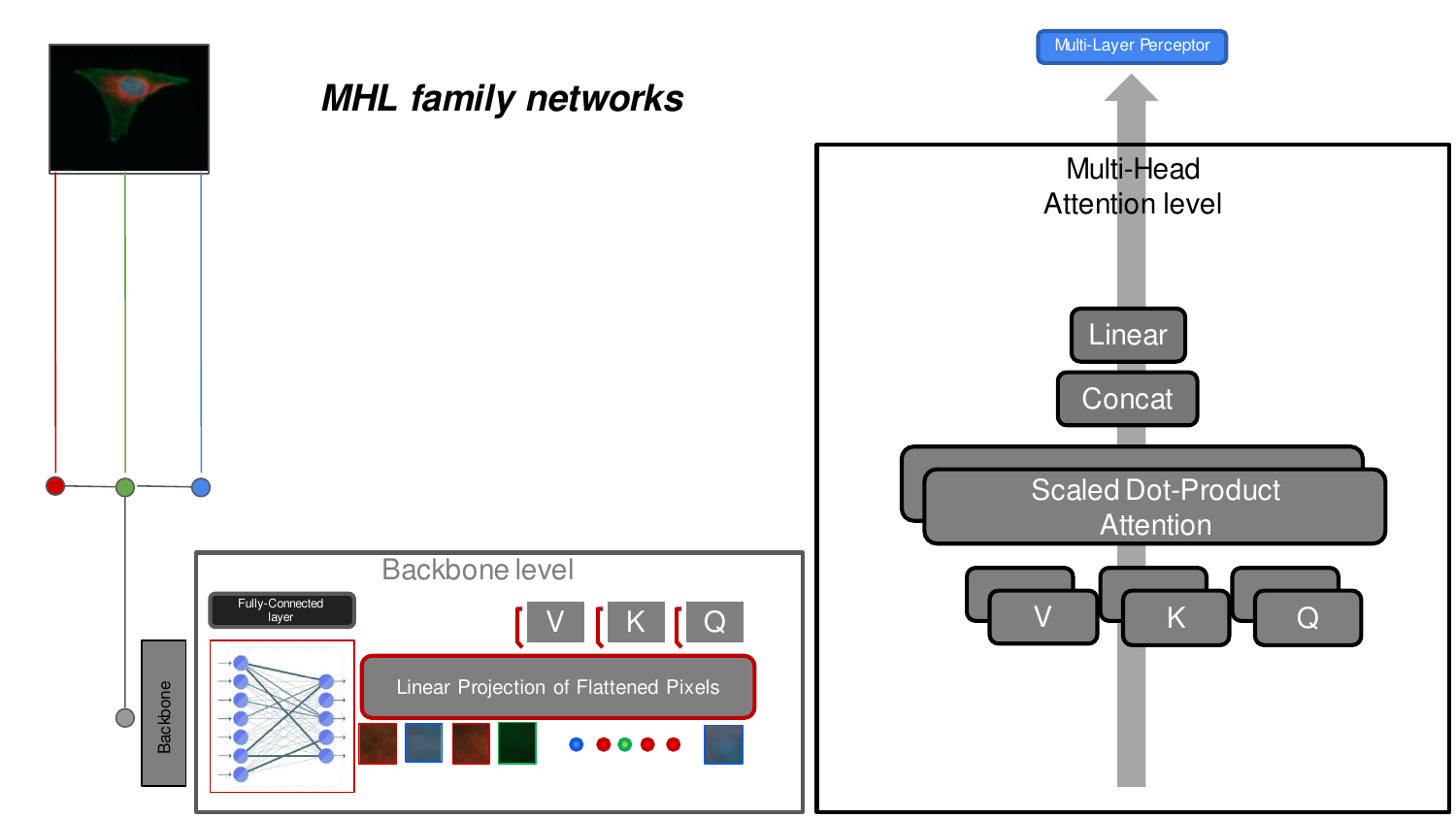}
\caption{The networks and the multi-head layer. The multihead layer (MHL) family (MHL, MHL-Res, MHL-Den).The MHL architecture has two main levels backbone, multi-head attention. All the three channels of the input are used as a single input in the backbone level and there is no isolated level like in RGB architecture. The Backbone level involves established classifiers (convolution networks or ResNet-50 or DenseNet-121) which use as input the RGB image. The Multi-head attention level uses the output from the backbone level to initialise a scaled dot-product self-attention layer which follows a concatenation and a multi-layer perceptor to classify the input image.}
\label{ma2}
\end{figure*}

Two different families of networks were developed; the RGB (red, green and blue) family (Fig. \ref{ma1}) and the {multi-head layer} (MHL) family (Fig. \ref{ma2}). The main differences of these two families of deep learning networks are that RGB family use an isolator level to isolate the three channels of the cell's images, contrary to the MHL family which uses all three channels in the backbone, and multi-head attention levels. Each family had three different networks (RGB, RGB-Res, RGB-Den and MHL, MHL-Res, MHL-Den) based on the different deep learning networks we used in the backbone level (Convolutional layer network, ResNet-50 and DenseNet-121).
Fig. \ref{ma1} and \ref{ma2} presents the different multi-head networks we used for this study. The RGB family encoder networks have three levels (Fig. \ref{ma1}). The first level is the isolator which isolates each of the three channels of the initial image. The next level is the backbone which uses a convolutional network (ResNet-50 or DenseNet-121, or a three-layer convolutional layer of one filter) three times for each isolated channel of the input image. The output of this level are three flatten 15000x2 signals (2 is the number of classes in the classification task). Each of these signals is encoded in value (V), key (K), and queries (Q) metrics as described in \citep{mha} study. The multi-head attention level uses each output from the backbone network (V, K, Q) to initialise a scaled dot-product attention layer. Six multi-head attention layers \citep{mha} are used (three self-attention of each channel (Red, Green, Blue) and three pair channels multi-attention layers (Red/Green, Blue/Green, Red/Blue)). The last layer in the multi-head attention is a multi-layer perceptron (MLP). MLP is a combination of two dense layer (1024, 512 of Relu activation) and one dense layer of the classification number (here is 2, of Softmax activation). The multi-layer perceptron has two dropout layers to estimate the epistemic uncertainty of the model and to reduce the overfitting effect of the model. On the other hand, MHL family encoder networks have no isolation level (Fig. \ref{ma2}). The networks use all the RGB channels of the colour image as input in the backbone level. The output of the backbone level is a flatten 65536x2 signal (2 is again the number of classes in the classification task). In the multi-head attention level the networks use a scaled dot product self-attention. We called it "self-attention" as we used twice the same output of the backbone level. The multi-layer perceptron layer is the same as in the RGB family.
The attention heads for the RGB family networks are two (Red/Blue, Red/Green, Blue/Green, Blue/Blue, Green/Green, and Red/Red: the six multi-head attention layers we used) and for the MHL family networks are two (self-attention, twice the same output of the backbone level).

The mathematical formalization of multi-attention level of RGB family  $MAL_{RGB}$ is given by:
\begin{equation}
MAL_{RGB}=Concat(D_{head})
\label{01}
\end{equation}
where $Concat$ is a concatenate filter.
$D_{head}$ is a symmetric square matrix given by:
\begin{equation}
D_{head}=
\begin{bmatrix}
dh(R,R) & dh(R,G) & dh(R,B) \\
dh(G,R) & dh(G,G) & dh(G,B) \\
dh(B,R) & dh(B,G) & dh(B,B) \\
\end{bmatrix}
\label{11}
\end{equation}
where R is the backbone output of the red channel, and B,G the backbone output of the blue and green channels respectively. In RGB family we used six elements of the $D_{head}$ matrix as we did not want to duplicate the information of the symmetric elements ($dh(G,R)=dh(R,G)$, $dh(G,B)=dh(B,G)$ and $dh(B,R)=dh(R,B)$). 
Thus the $D_{head}$ of equation \ref{11} becomes an upper triangular matrix given by:
\begin{equation}
D_{head}^{upper}=
\begin{bmatrix}
dh(R,R) & dh(R,G) & dh(R,B) \\
0 & dh(G,G) & dh(G,B) \\
0 & 0 & dh(B,B) 
\end{bmatrix}
\label{111}
\end{equation}
Each element of the upper triangular matrix was given by:
\begin{equation}
dh(i,j)=MHA_{n=2}(Q,K,V)
\label{121}
\end{equation}
where $i, j$ $\in {[R,G,B]}$. if $i=j$ then we have a self-attention layer. $MHA_{n=2}$ is a multi-head attention layer of two heads ($n=2$). The $MHA_n$ is given by:
\begin{equation}
MHA_{n=N} (Q,K,V) = Concat(head_{c=1}, ..., head_{c=N})W^O
\label{13}
\end{equation}
where $N$ is the number of the different heads (in our study the three backbone output channels R,G and B) and $W^O$ weights were described analytically in \citep{mha}.
The $head_c$ is given by:
\begin{equation}
head_c=AT(QW^Q_c,KW^K_c, VW^V_c)
\label{14}
\end{equation}
where $c$ is the backbone output channel number (the backbone can be either R,G or B). The $AT$ is the attention layer and it computed by:
\begin{equation}
AT(Q,K,V)= softmax(\frac{QK^T}{\sqrt{d_k}})V
\label{15}
\end{equation}
where the input matrix are combination of queries and keys of dimension $d_k$, and values of dimension $d_v$. Queries packed together into matrix $Q$. The keys and values are also packed together into matrices $K$ and $V$. 

Based on equations \ref{01} the output of the MHL family networks is given from:

\begin{equation}
MAL_{MHL}=Concat(D_{head})
\label{22}
\end{equation}
where the $D_{head}$ matrix of equation \ref{11} becomes:
\begin{equation}
D_{head}=dh(RGB,RGB)
\label{23}
\end{equation}
where $RGB$ is the backbone output signal of all the channels (Red, Green, Blue) and the $dh(i,j)$ of equation \ref{121} is a self-attention layer (Fig. \ref{ma2}).

\subsection{Analyzing networks pattern learning through a combination of local Explainable and global interpretable techniques}
Interpretability and explainability is a very important part of a classification study as it verifies the correct training of the machine learning network. In this study we used the most common local explainability sensitivity technique in the medical imaging applications, the GradCam method \citep{xai22}. To compute the class-discrimative localization map of the width $w$ and the height $h$ of a specific cell image for a class $c$ (metastatic or healthy), we compute the gradient of the score for the class $c$, $y^c$ with respect of the $k^{th}$ feature activation map ($A^k$) of the last convolution layer in each deep network. To compute the weights of the importance ($\alpha^c_k$) of each $k$ feature activation map, we used a global average pooling over the width ($i$) and height ($j$) of each feature:
\begin{equation}
\alpha^c_k=\frac{1}{Z}\sum_i{{\sum_j\frac{dy^c}{dA_{ij}^k}}}
\label{20}
\end{equation}
where $Z$ the summation of $i$ and $j$.
Moreover, we used a weighted combination of forward activation maps and a $ReLU$ to deliver the final GradCam activation map.
\begin{equation}
GradCam=ReLU(\sum_k{\alpha^c_k A_{ij}^k})
\label{21}
\end{equation}
Furthermore, as we wanted to globaly validate and study the pattern learning of the deep networks, we used an interpretable technique to generalize the local explanations of GradCam. Thus, we computed a weighted geometric mean activation map from all the cell images (Cells Gmean-Shape) and their local GradCam results (Gmean-GradCam). The weighted geometric mean was given by:

\begin{equation}
Gmean(X,W)= exp(\frac{\sum{w_i lnx_i}}{\sum{w_i}})
\label{12}
\end{equation}

where W is the weight tensor and the X is the cells and GradCam images tensor. The weight value of the weight tensor for each image can be between 1.0 to 0.0. In our implementation we use 1.0 for all the images.
To evaluate the learning patterns of the networks we computed a statistical correlation coefficient image between the Gmean-Shape and Gmean-GradCam images. We utilized a patch size ($p_1,p_2$) to compute each of the correspond pixels of the correlation coefficient image. After trial the patch size was chosen 2x2 pixels size. The statistical correlation coefficient equation was given by: 
\begin{equation}
C_{coef}(p_1,p_2)=\frac{\frac{1}{N}\sum_{n=0}^N([p_1-\overline{p_1}][p_2-\overline{p_2}])}{\sigma_{p_1}\sigma_{p_2}}
\label{221}
\end{equation}
where $\sigma_{p_1}$ and $\sigma_{p_2}$ the standard deviation of the patch of Gmean-Shape ($p_1$) and Gmean-GradCam ($p_2$) images respectively. N is the number of pixel in each patch. 
Each pixel of the correlation coefficient image was compute based on the equation:
\begin{equation}
C(i,j)=C_{coef}(Gm^{S}[i\pm2,j\pm2],Gm^{GC}[i\pm2,j\pm2])
\label{231}
\end{equation}
where $Gm^{S}$, $Gm^{GC}$ are the GmeanShape, GmeanGradCam images respectively and the patch size is $[i\pm2,j\pm2]$.
Lastly, a positive and negative ratio score of all the pixels of the correlation coefficient image was extracted so we can study the relation of the developed explainable technique and the shape variation of the human cells.

All the continuous variables are presented as proportions, means ± standard deviations, or median and interquartile range for data not following a normal distribution. Variable standardisation was performed to allow comparison of the different continuous variables on the same scale by subtracting the mean for each variable and dividing it by its standard deviation (SD). Normally distributed variables were correlated using Pearson’s correlation coefficient, otherwise, the Spearman correlation coefficient was used. T-Test analysis used to calculate the statistically significant differences between the different networks we used.

\section{Experiments}
\subsection{Experiment details}
To evaluate our methodology, we demonstrate our approach on a dataset experiment composed on fluoresence microscopy images of cancer and normal cells. The evaluation task is binary-class classification between normal and metastasizing cells. As normal cells we use Bj primary fibroblast cells. As metastasizing cells, we used the isogenically matched, oncogenically transformed counterpart BjTertSV40TRasV12, which forms invasive tumours when injected in immunodeficient mice \citep{cell}. Bj Primary and BjhTertSV40T cells, previously created by Hahn et al., \citep{cell} were cultured in Dulbecco’s modified Eagle’s medium supplemented with 10\% foetal bovine serum and 1\% penicillin/streptomycin, at 37 $ ^\circ C $ and 5\% $CO_{2}$, as described previously \citep{10}. Cells were seeded at a density of 4000/$cm^2$ (BjhTertSV40T) or 8000/$cm^2$ (Bj Primary) on glass coverslips to obtain a 40\% confluent monolayer of cells after 48 hours. Post 48h, the cells were fixed in 4\% paraformaldehyde for 15 minutes at 37 $ ^\circ C $, followed by permeabilisation in 0.2\% Triton-X100 for 3 minutes at room temperature and blocking in 1\% Bovine serum albumin (Sigma, St Louis, MO, USA) for 1 hour at room temperature. The cells were then immunostained as described previously \citep{11}. We stained the nuclei with 4',6-diamidino-2-phenylindole (DAPI), a blue fluorescence stain that binds to double-stranded DNA. It has an absorption maxima of 358 nm and emission maxima of 461 nm. The actin cytoskeleton was stained using phalloidin, a peptide with a high binding affinity for filamentous actin. Phalloidin is labelled with a green fluorescent probe $Alexa Fluor^{TM}$ 488 at 0.0132 $\mu$M (A12379, Thermo Fisher, Waltham, MA, USA). It has an absorption maxima of 490 nm and emission maxima of 525 nm. Vimentin was stained using a mouse anti-vimentin antibody at 3.3 $\mu$g/mL (V6389, Sigma, St Louis, MO, USA), which is then visualised using an anti-mouse secondary antibody conjugated to the red fluorescent probe $Alexa Fluor^{TM}$ 647 goat anti-mouse Alexa 647 at 5 $\mu$g/mL (A-21235, Thermo Fisher, Waltham, MA, USA). It has an absorption maxima of 650 nm and emission maxima of 671 nm. Coverslips were mounted on glass slides with Mowiol (Sigma, St Louis, MO, USA). Imaging was performed with a Zeiss CellDiscoverer 7 Microscope and Zeiss ZEN software (Zeiss, Oberkochen, Germany) using a 20X objective. 10x10 tiled images were taken with 10\% overlap. Identical exposure time and LED intensity were used for all images.
\subsection{Cohort's prepossessing image analysis}
Image analysis techniques have been applied to all slices to reduce the artefacts. We have used noise filters such as binomial deconvolution, Landweber deconvolution \citep{Vonesch2008}, and curvature anisotropic diffusion image filters \citep{Perona1990} to reduce noise in the images. Moreover, all images were re-sampled to have dimensions 512 to 512 pixels. The initial cohort was a combination of 100 normal cell and 120 oncogenically transformed, invasive, metastasizing cells of three channels (RGB) images. The validation protocol was a combination of partial offline data augmentation and a 5-fold stratified cross-validation scheme. The data augmentation techniques included rotation (rotation around the centre of the image by a random angle in the range of $-15^{\circ} $ to $ 15^{\circ}$), width shift range (width shift of the image by up to $20$~pixels), height shift range (height shift of image by up to $20$~pixels), and ZCA whitening (add noise in each image) \citep{ko}. 
\subsection{Hyper-parameters initialisation and networks training}
After random shuffling, we used a 5-fold structural cross validation scheme. In each of the five folds we used new generated augmented images of the normal and metastasizing cells (352, two times the data of the 4 training folds, partial offline data augmentation technique). In each of the five folds we used 44 (1 fold without data augmentation) 'unseen' images for testing. We have used a binary cross-entropy as the cost function. The loss function is optimised using the stochastic gradient descent (SGD) method with a fixed learning rate of $0.001$. We have applied transfer learning techniques to the VGG-16, DenseNet-121, ResNet-50 and DensRes-131 networks using the ImageNet dataset \citep{imagenet} (\url{http://www.image-net.org}). The RGB and MHL networks used as backbone pre-trained ResNet-50, DenseNet-121 and CNN networks for 100 epochs. The validation protocol was the same for the pre-trained backbone networks and the RGB, MHL networks so we can avoid any data leak in the training and validation experiment.  All networks were trained in two regimes: (a) less than 100 epochs and (b) more than 100 epochs respectively, to study the sensitivity of the networks' performance with respect of the training iterations. 
\subsection{Code and Data availability}
 The code used in our experiment was developed in a NVIDIA cluster (JADE2) with $4$ GPUs and $64$~GB RAM memory. The code is publicly available in \url{https://github.com/ece7048/MHL-RGB-deep-learing}. The dataset and the trained deep learning models can be provided after request to the corresponding authors.
 
\section{Results}
\begin{figure*}[h] 
\centering{
\includegraphics[trim={2.5cm 0.45cm 2.5cm 0.45cm}, scale=0.7]{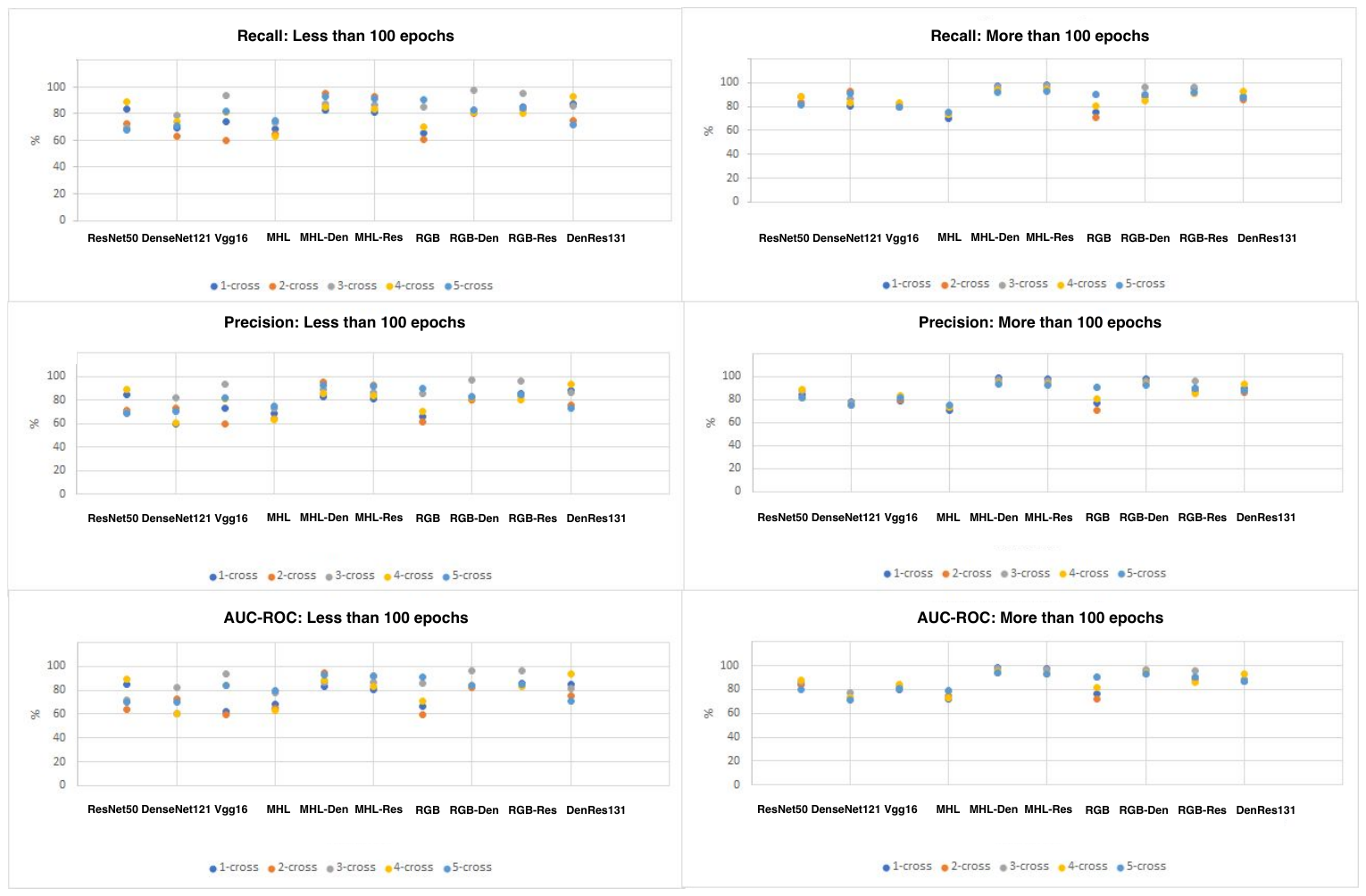}}
\caption{The Recall, Precision and AUC-ROC metrics results of the DL, MHL and RGB families in less than 100 epochs and more than 100 epochs of the 5-fold cross-validation scheme. The DL family is the ResNet-50, Densenet-121, Vgg-16 and DenRes-131. The MHL and RGB families are MHL, MHL-Res, MHL-Den and RGB, RGB-Res, RGB-Den respectively.}
\label{metric1}
\end{figure*}

We developed a new multi-head attention deep learning networks and a combination of local explainable and global interpretable techniques. To evaluate and verify the performance of the different networks we extract ROC-curves, for a number of metric scores (f1, recall, precision, AUC-ROC), and box-blots to capture their accuracy and robustness. Tables \ref{tab}, \ref{tab2} summarize the metrics results of the different established deep learning (Vgg-16,ResNet-50, DenseNet-121, and DenRes-131), and the attention deep learning networks (ViT, RGB family (RGB, RGB-Res, RGB-Den) and MHL family (MHL, MHL-Res, MHL-Den)). We split the metrics results in two cases' the $< 100$ epochs, and the $> 100$ epochs. Based on the Figure \ref{metric1} and the Tables \ref{tab}, \ref{tab2}, it was observed that in the smaller number of epochs all the families of networks were less robust with standard deviations $> 6$ \% than that found for networks trained with the larger number of epochs ($< 3$ \%). In addition all the networks showed better performance (metrics score) from the $> 100$ epochs compared to the $< 100$ epochs case.

MHL-Den, RGB-Den, and DenRes-131 deliver the highest performance in all the metrics. More thoroughly, about the Recall, Precision, AUC-ROC and f1-score: DenRes-131 performed 88.24, 88.64, $>86.00$, and $>88.00$, MHL-Den outperformed 94.89, 94.90, $>94.90$ and $>94.00$ respectively. Lastly, RGB-Den performed 93.73, 93.77, $>94.00$, and $>93.80$ respectively. The most robust network based on the metrics standard deviation were the RGB-Res, RGB-Den, and DenRes-131 with less than 0.8, 1.8 and 1.7 \% in all the metrics.   

\begin{figure}[h!]
\caption{The ROC-curves for the established deep learning networks (DL family). With blue line the healthy cells class (1 class) with red line the metastatic cells class (2 class)}
\label{r1}
      \medskip    
\centerline{
\relax \textbf{(a)}
    \includegraphics[trim={0.00cm 0.cm 0.0cm 0.00cm},clip,scale=.32]{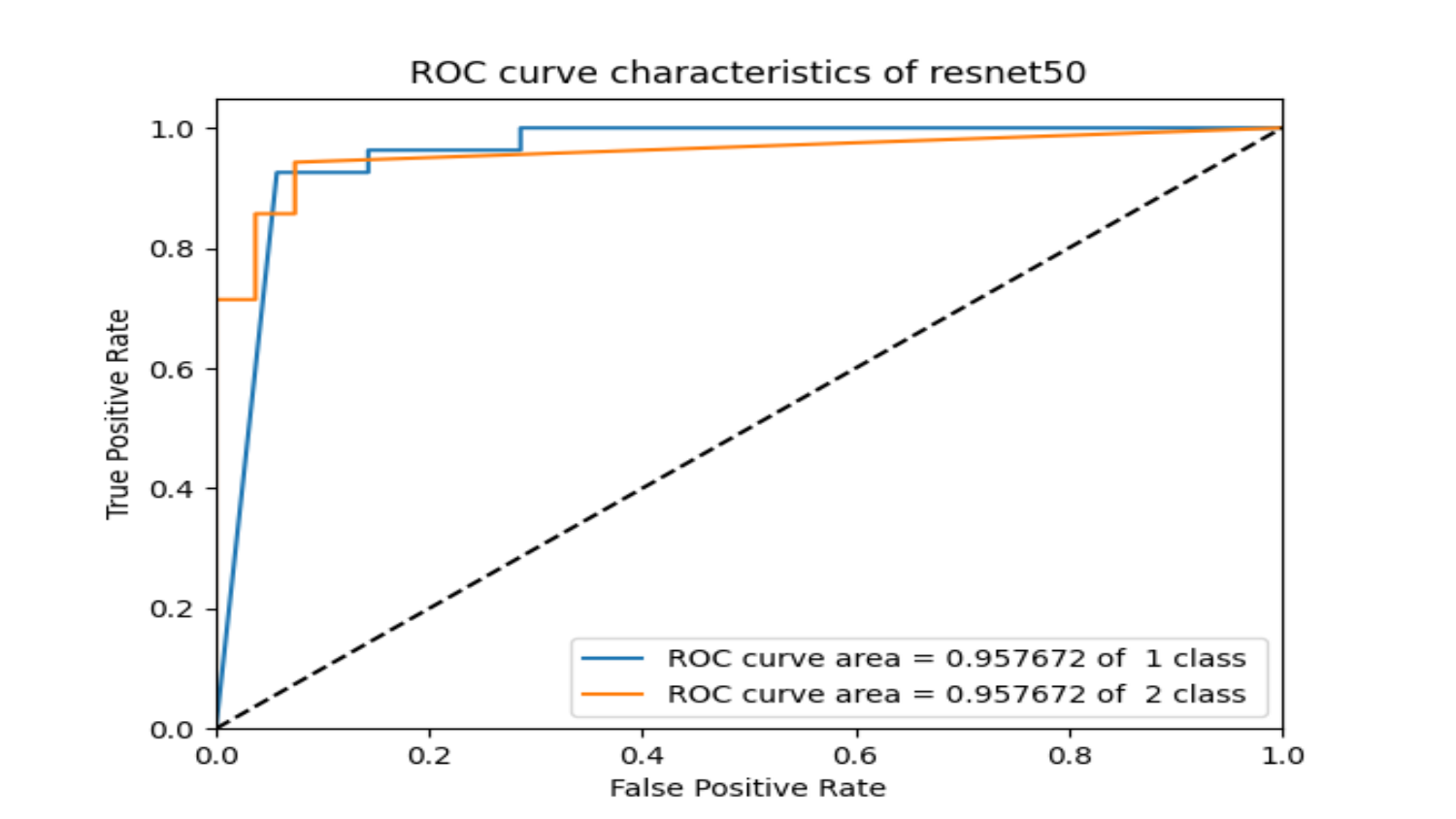}
\relax \textbf{(b)}
    \includegraphics[trim={0.6cm 0.0cm 0.0cm 0.00cm},clip,scale=.35]{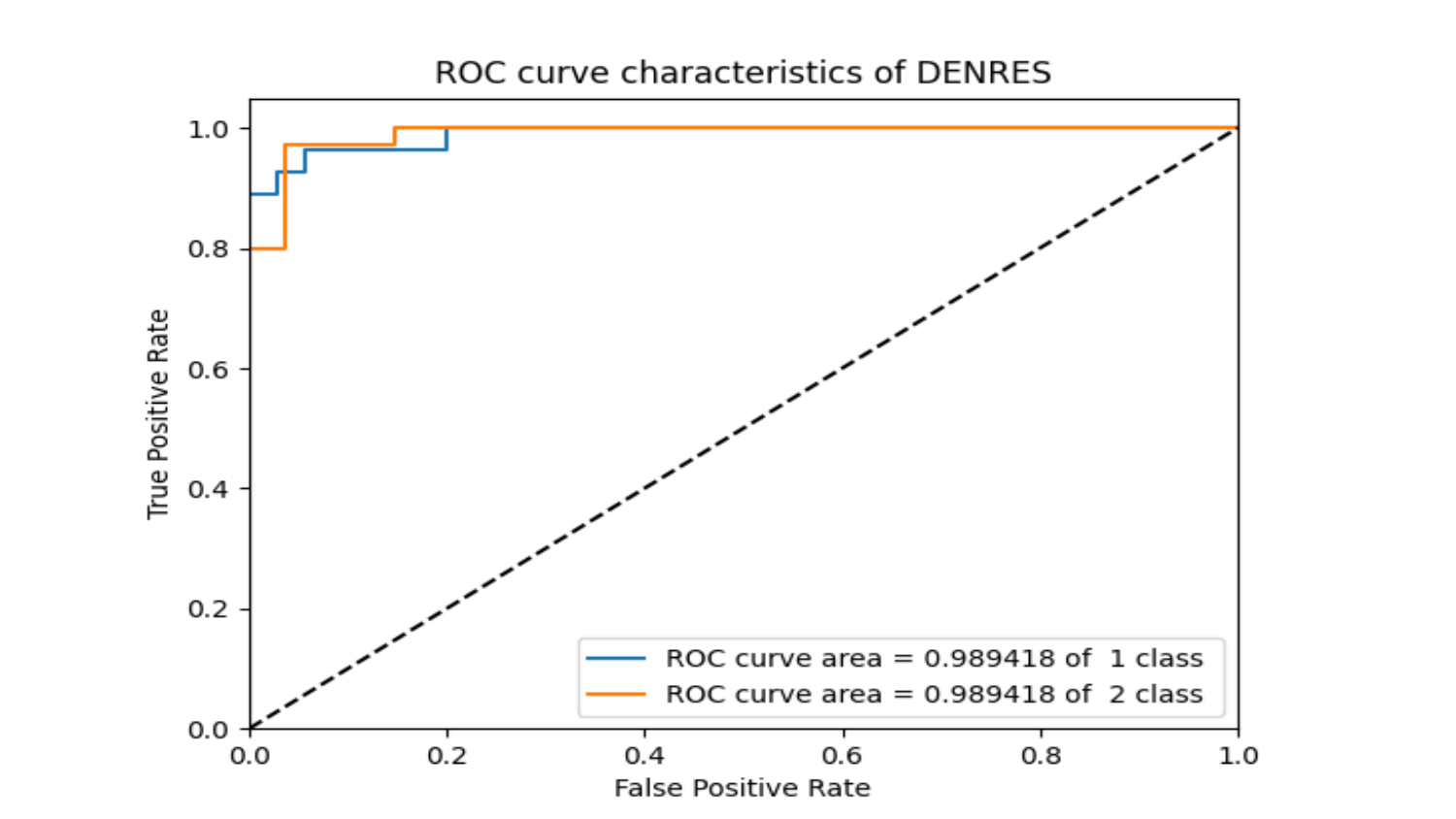}}
\centerline{
\relax \textbf{(c)}
    \includegraphics[trim={0.60cm 0.cm 0.00cm 0.00cm},clip,scale=.35]{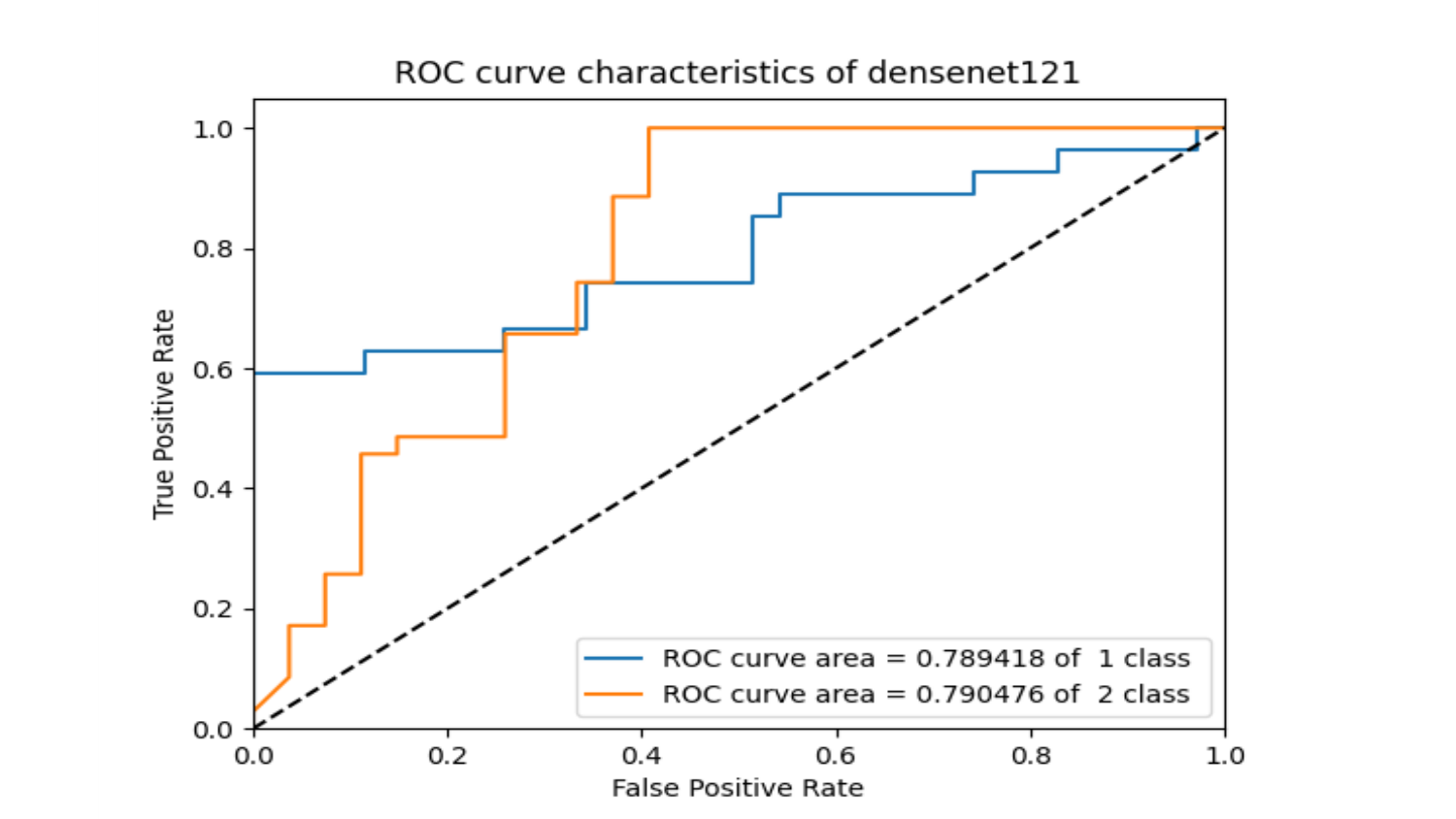}
\relax \textbf{(d)}
    \includegraphics[trim={0.60cm 0.cm 0.00cm 0.00cm},clip,scale=.35]{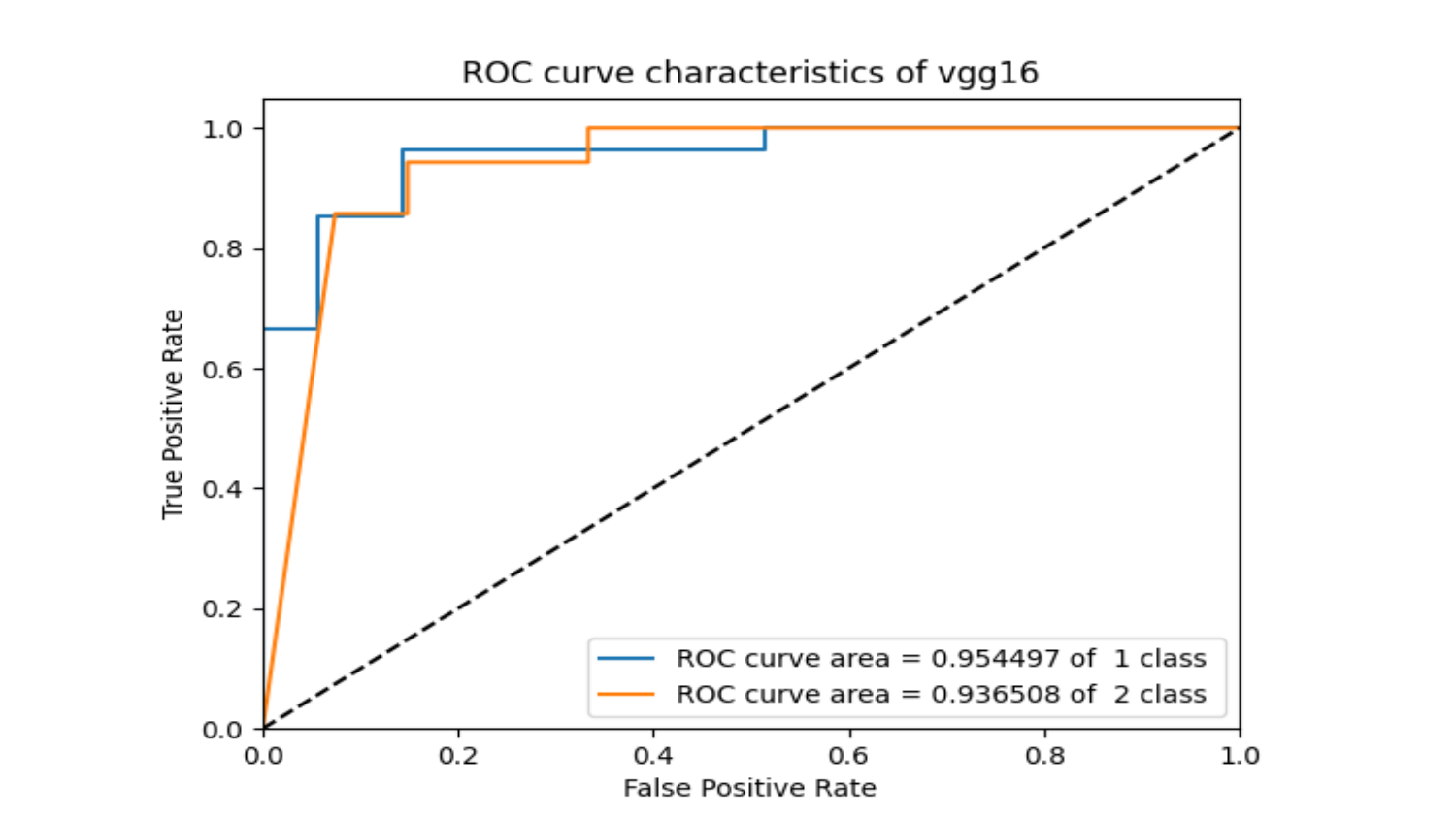}}

      \medskip    
\end{figure}

\begin{figure*}
\caption{The ROC-curves for the multi-head encoder networks (RGB and MHL families). With blue line the healthy cells class (1 class) with red line the metastatic cells class (2 class).}
\label{r2}
      \medskip    
\centerline{
\relax \textbf{(a)}
    \includegraphics[trim={0.00cm 0.cm 0.0cm 0.00cm},clip,scale=.36]{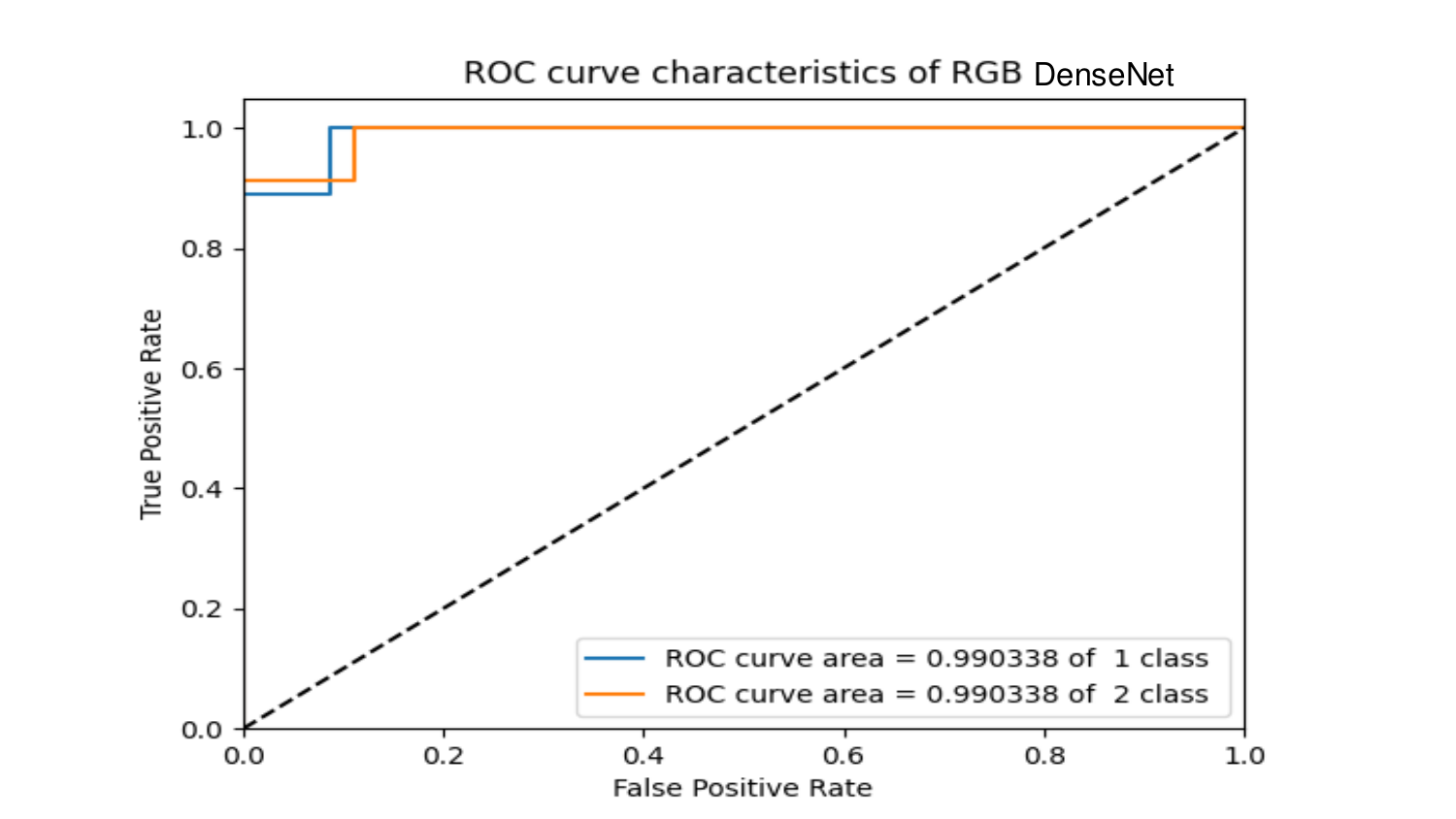}
\relax \textbf{(b)}
    \includegraphics[trim={0.6cm 0.0cm 0.0cm 0.00cm},clip,scale=.385]{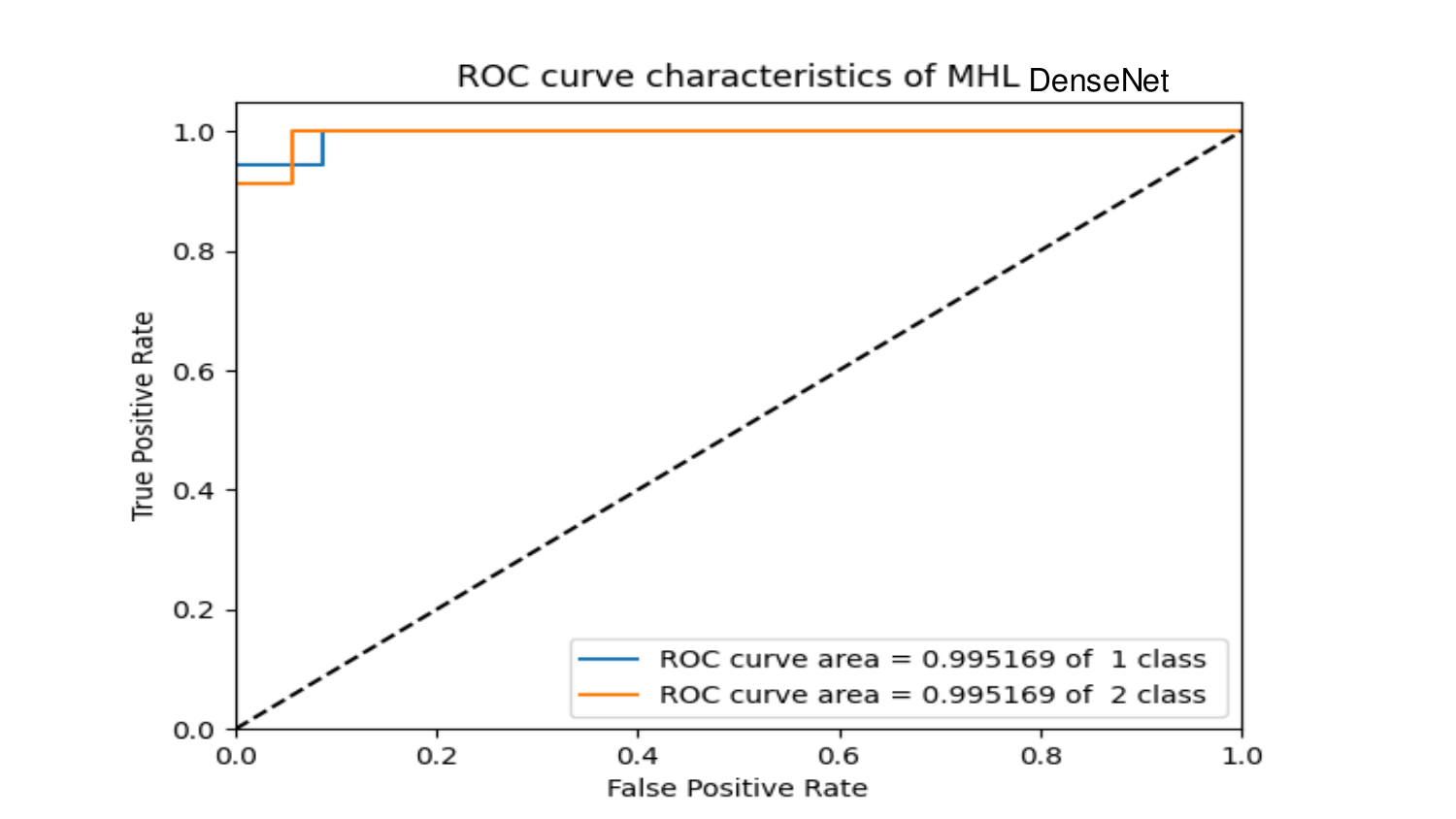}}
\centerline{
\relax \textbf{(c)}
    \includegraphics[trim={0.60cm 0.cm 0.00cm 0.00cm},clip,scale=.385]{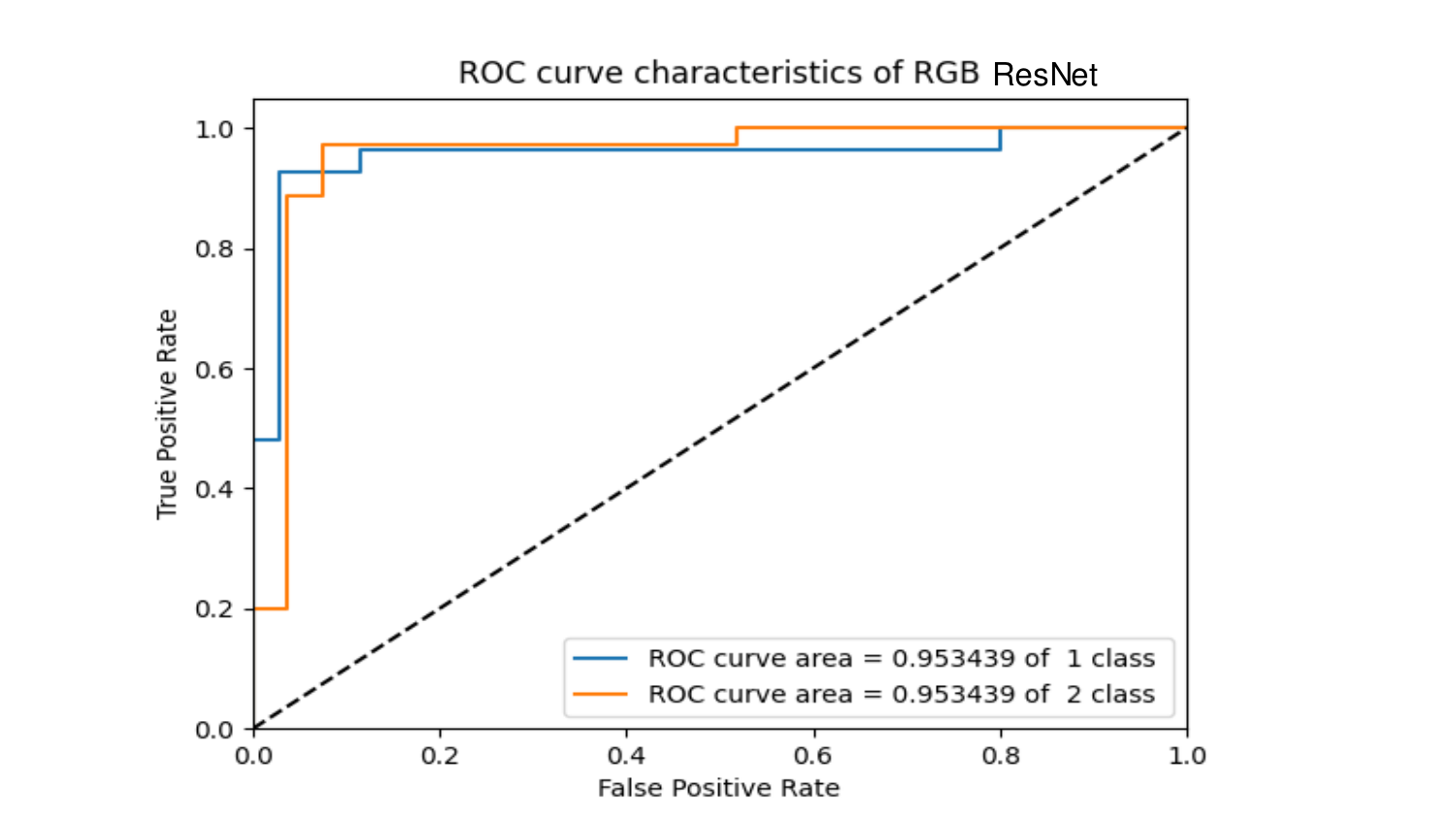}
\relax \textbf{(d)}
    \includegraphics[trim={0.60cm 0.cm 0.00cm 0.00cm},clip,scale=.385]{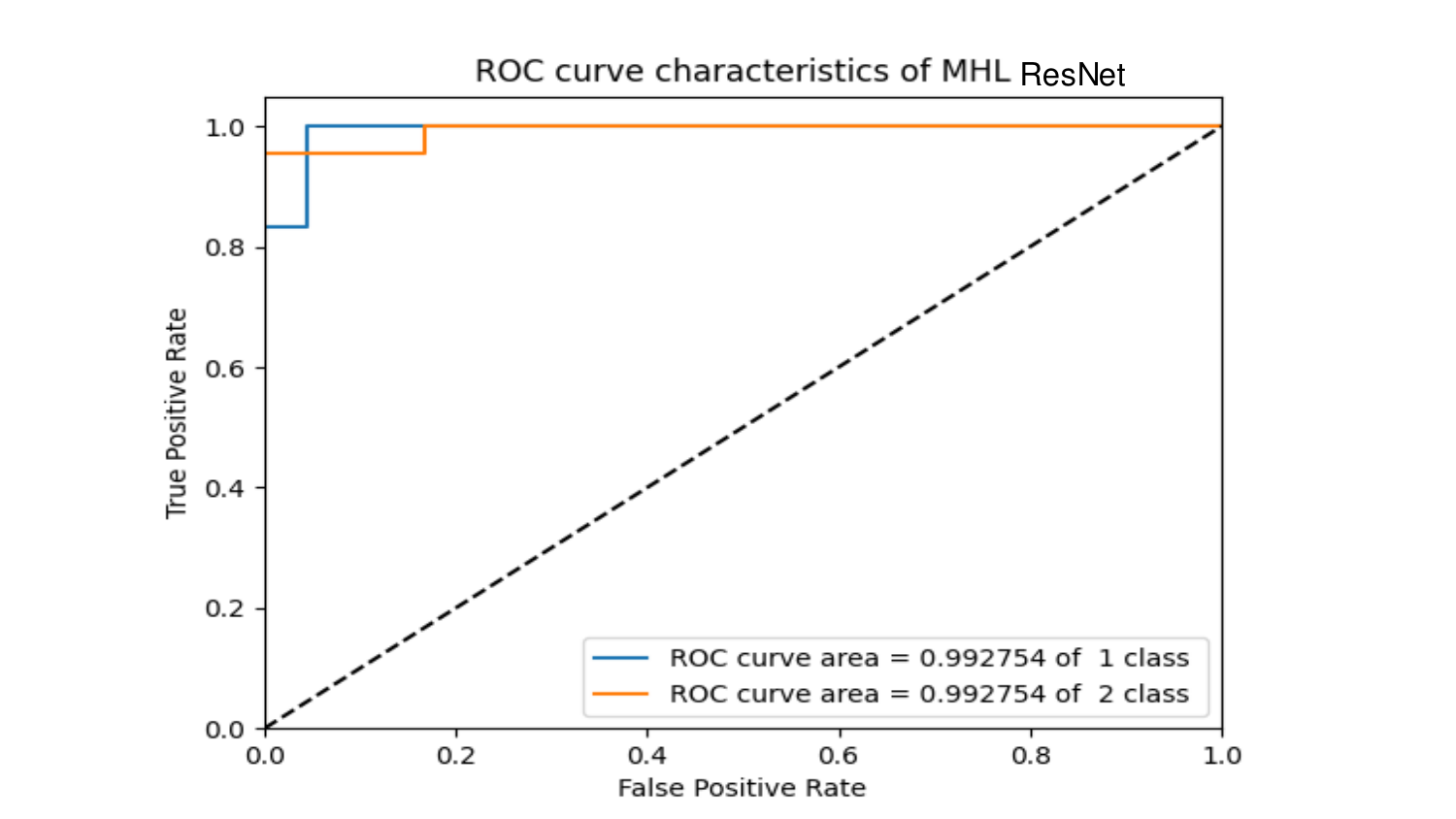}}
    
\centerline{
\relax \textbf{(e)}
    \includegraphics[trim={0.60cm 0.cm 0.00cm 0.00cm},clip,scale=.385]{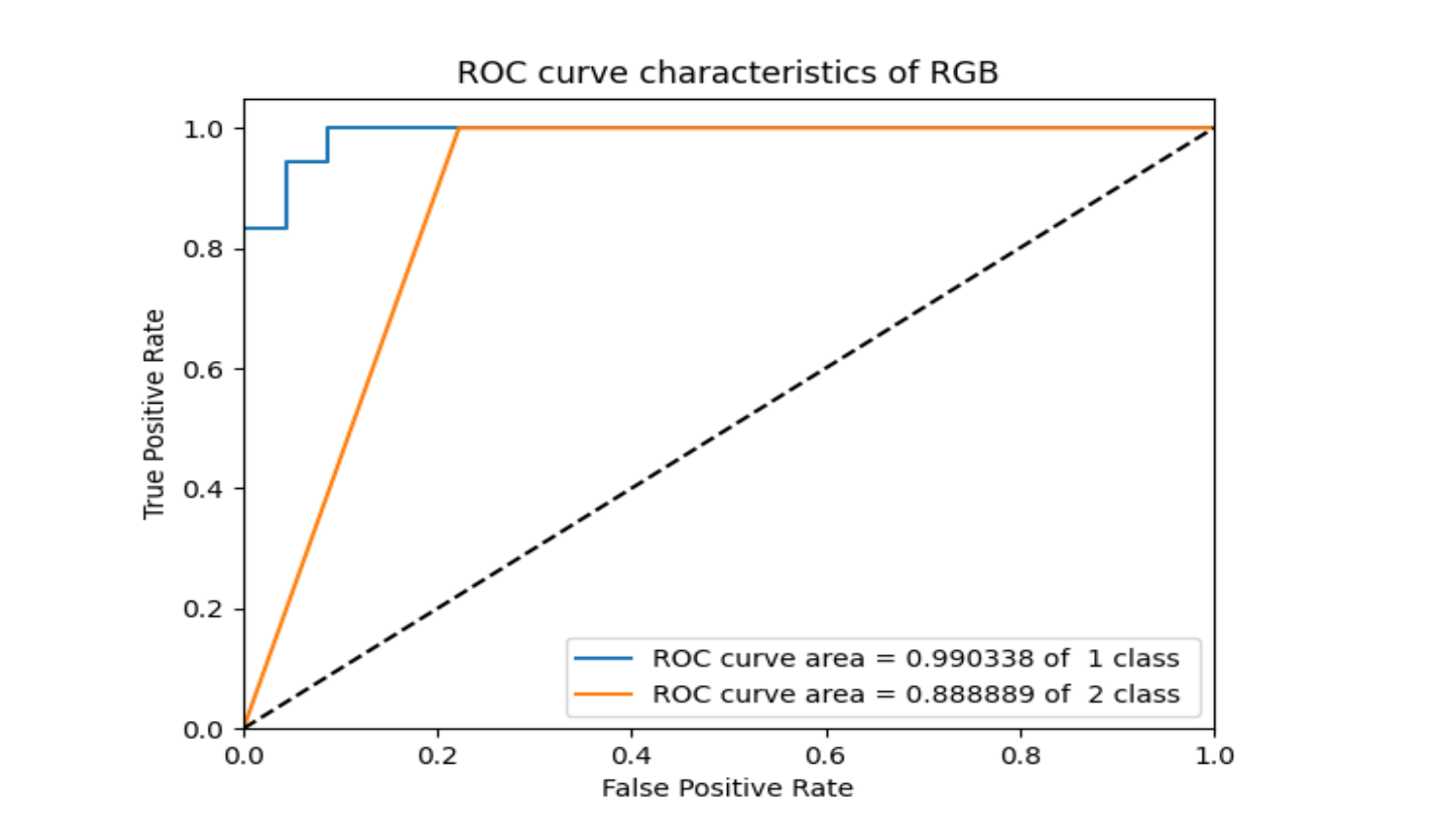}
    \relax \textbf{(f)}
    \includegraphics[trim={0.60cm 0.cm 0.00cm 0.00cm},clip,scale=.385]{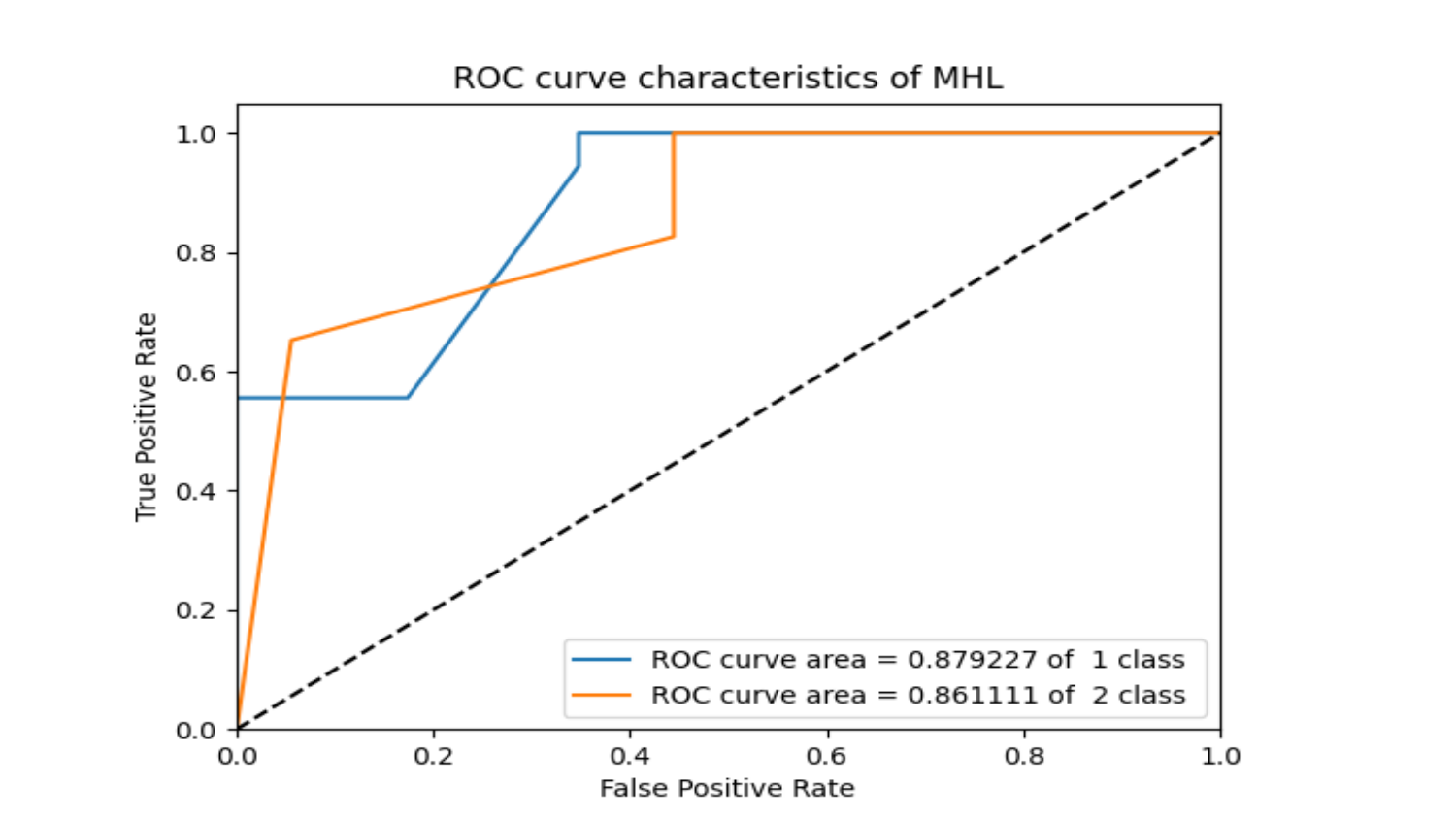}}
        \medskip      

\end{figure*}

Fig. \ref{r1}, \ref{r2} show the ROC curve of the three different family of networks (DL, RGB, MHL). The highest true positive rate in classes 1 and 2 (healthy and metastasizing cells) of the DL family (Fig. \ref{r1}) was in DenRes-131 (98.94, 98.94 \%) and the worst in the DenseNet-121 (78.94, 79.04 \%). Moreover, in the RGB family (Fig. \ref{r2}) RGB-Den outperformed the other networks with 99.04, 99.03 \%, and the RGB delivered the worst performance (99.04, 88.88 \%). Lastly in the MHL family (Fig. \ref{r2}), MHL-Den achieved the best ROC-curve performance with 99.51 and 99.52 \% in the normal and metastasizing cells classes respectively, and the worst the MHL network (87.92, 86.11\%). 

\begin{figure}
\centering 
\centering
\includegraphics[scale=.35]{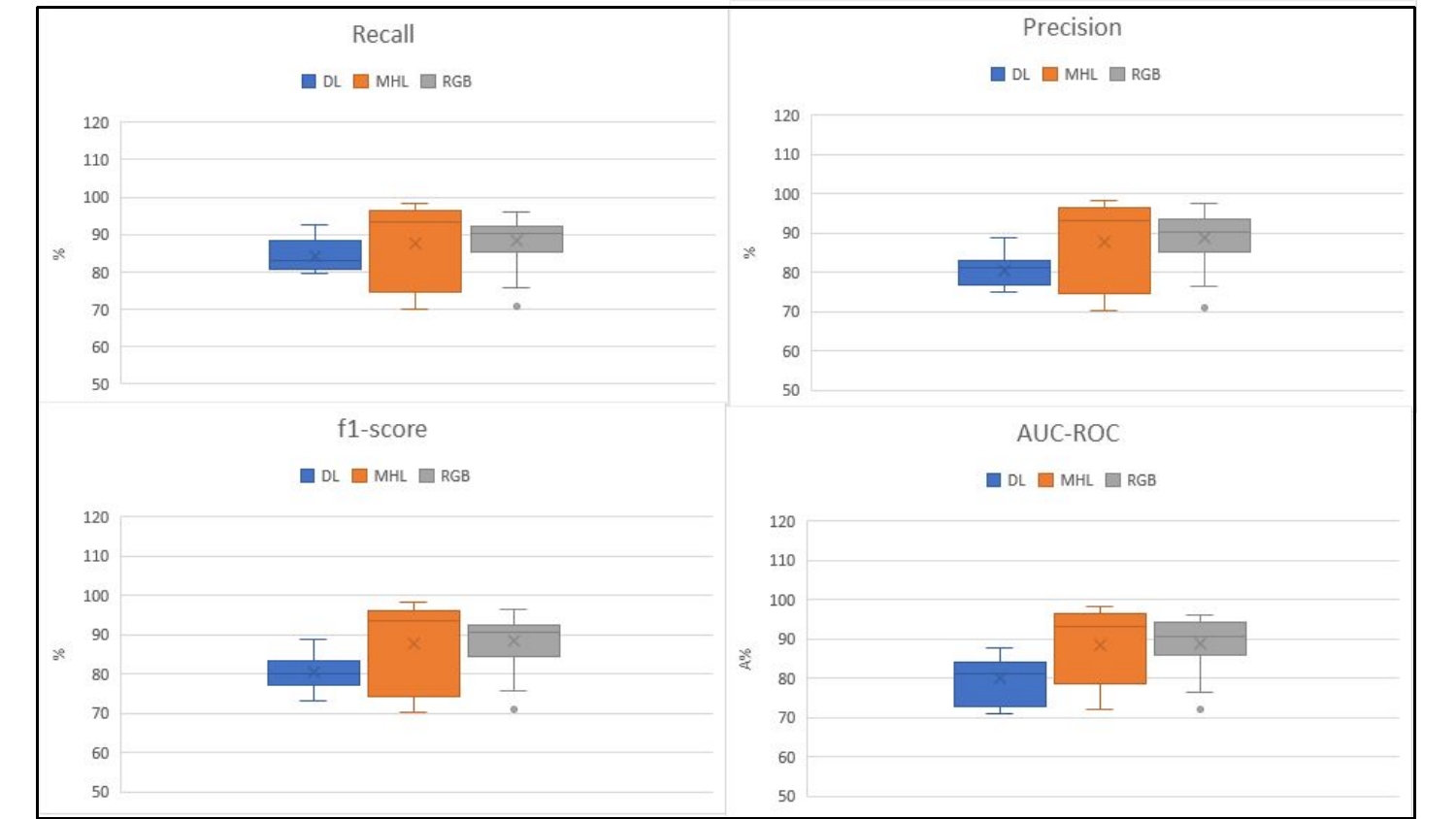}
\caption{Box-plots show Precision, Recall and f1-score metrics variability and robustness of the three families. The DL family is the ResNet-50, Densenet-121, Vgg-16 and DenRes-131. The MHL and RGB families are MHL, MHL-Res, MHL-Den and RGB, RGB-Res, RGB-Den respectively}
\label{metric2}
\end{figure}

Fig. \ref{metric2}, \ref{metric1} summarises the robustness of the networks in a variation of epoch training. The MHL families are the less robust networks compared with the DL and RGB families. RGB family were more robust in the AUC-ROC and Recall metrics and the DL family was more robust in the f1-score and Precision metrics. However, even if the RGB and DL networks were more robust (small standard deviation) MHL outperformed both in the average score of each of the four metrics.  

\begin{table*}
\caption{Quantitative evaluation metrics on the cells Sheffield dataset. The table shows the average values and standard deviation ($\pm$) of a 5-fold structural cross-validation scheme.}
\centering

\begin{tabular}{ |p{2.6cm}|p{1.95cm}|p{1.95cm}|p{1.8cm}|p{1.8cm}|}
\hline
\multicolumn{5}{|c|}{Binary classification performance of DL networks in $< 100$ epochs} \\
\hline
Metric  (\%)  & Vgg16 & Resnet50 & DenseNet121 & DenRes131 \\
\hline
Recall  & 78.35 $\pm$ $5.52$ & 76.44 $\pm$ $9.33$ & 70.27 $\pm$ $1.31$ & 82.74 $\pm$ $8.72$ \\
Precision   & 77.75 $\pm$ $6.32$ & 76.36 $\pm$ $9.44$ & 71.23 $\pm$ $2.61$& 83.34 $\pm$ $8.72$\\
AUC-ROC macro   & 73.15  $\pm$ $15.83$ & 75.55 $\pm$ $10.02$ & 65.10 $\pm$ $7.03$ & 81.10 $\pm$ $8.65$ \\
AUC-ROC micro   & 77.25 $\pm$ $7.08$ & 76.78 $\pm$ $8.93$ & 71.23  $\pm$ $2.62$ & 82.60 $\pm$ $9.11$ \\
AUC-ROC weight  & 73.14 $\pm$ $15.72$ & 77.48 $\pm$ $8.63$ & 65.66 $\pm$ $7.84$ & 81.24  $\pm$ $8.25$ \\
F1  sample  & 77.80 $\pm$ $6.28$ & 76.68 $\pm$ $9.06$ & 70.30 $\pm$ $1.35$  & 84.47 $\pm$ $8.39$ \\
F1 macro   & 64.55 $\pm$ $23.42$ & 70.13 $\pm$ $11.13$ & 55.43 $\pm$ $6.31$ & 83.50 $\pm$ $8.30$\\
F1 micro    & 77.30 $\pm$ $7.02$ & 76.74 $\pm$ $8.94$ & 70.72 $\pm$ $2.05$ & 84.36 $\pm$ $8.57$ \\
F1 weight   & 73.26 $\pm$ $13.34$ & 74.54 $\pm$ $11.49$ & 63.36 $\pm$ $9.26$ & 83.69 $\pm$ $8.92$ \\
\hline
\multicolumn{5}{|c|}{Binary classification performance of DL networks in $> 100$ epochs} \\
\hline
Metric  (\%)  & Vgg16 & Resnet50 & DenseNet121 & DenRes131 \\
\hline
Recall  & 79.75 $\pm$ $0.42$ & 82.24 $\pm$ $1.50$ & 75.28 $\pm$ $4.31$ & 88.24 $\pm$ $0.74$ \\
Precision   & 80.09 $\pm$ $1.65$ & 82.68 $\pm$ $2.02$ & 76.72 $\pm$ $2.22$& 88.64 $\pm$ $1.09$\\
AUC-ROC macro   & 80.54  $\pm$ $0.83$ & 82.12 $\pm$ $2.91$ & 72.02 $\pm$ $1.29$ & 86.97 $\pm$ $0.22$ \\
AUC-ROC micro   & 80.12 $\pm$ $1.41$ & 81.70 $\pm$ $2.78$ & 74.22  $\pm$ $2.81$ & 87.17 $\pm$ $0.29$ \\
AUC-ROC weight  & 80.56 $\pm$ $0.68$ & 82.53 $\pm$ $2.42$ & 71.68 $\pm$ $0.83$ & 86.88  $\pm$ $0.21$ \\
F1  sample  & 80.53 $\pm$ $0.79$ & 81.59 $\pm$ $2.51$ & 75.79 $\pm$ $3.65$  & 88.75 $\pm$ $1.64$ \\
F1 macro   & 80.55 $\pm$ $0.78$ & 81.55 $\pm$ $2.46$ & 76.42 $\pm$ $3.53$ & 88.02 $\pm$ $0.99$\\
F1 micro    & 80.70 $\pm$ $0.57$ & 81.65 $\pm$ $2.33$ & 76.25 $\pm$ $4.04$ & 88.78 $\pm$ $1.72$ \\
F1 weight   & 80.96 $\pm$ $0.35$ & 81.96 $\pm$ $3.13$ & 75.97 $\pm$ $4.06$ & 88.49 $\pm$ $1.28$ \\
\hline
\end{tabular}
  \label{tab}
\end{table*}

\begin{table*}
\caption{Quantitative evaluation metrics on the cells Sheffield dataset. The table shows the average values and standard deviation ($\pm$) of the 5-fold structural cross validation scheme.}
\begin{tabular}{ |p{2.6cm}|p{1.97cm}|p{1.97cm}|p{1.77cm}|p{1.78cm}|p{1.78cm}|p{1.78cm}|p{1.78cm}|}
\hline
\multicolumn{8}{|c|}{Binary classification performance of Attention DL networks in $< 100$ epochs} \\
\hline
 Metric  (\%)  & ViT & RGB & RGB-Res & RGB-Den & MHL & MHL-Res & MHL-Den \\
\hline
Recall   & 67.98 $\pm$ $16.11$ & 77.78 $\pm$ $16.42$ &84.72 $\pm$ $1.81$& 82.97 $\pm$ $1.76$ & 71.65 $\pm$ $4.83$& 86.05 $\pm$ $7.24$& 87.80 $\pm$ $6.85$\\
Precision    & 74.58 $\pm$ $16.02$ &77.84 $\pm$ $16.69$ &84.78 $\pm$ $1.97$& 82.98 $\pm$ $1.75$ &71.65 $\pm$ $4.84$& 86.11 $\pm$ $7.28$& 87.84 $\pm$ $6.80$ \\
AUC-ROC macro   & 75.83 $\pm$ $16.22$ &78.50 $\pm$ $16.51$&85.61$\pm$ $1.54$ & 83.53 $\pm$ $1.93$ & 73.93 $\pm$ $7.52$& 86.40 $\pm$ $7.71$& 87.86 $\pm$ $7.06$ \\
AUC-ROC micro   & 75.78 $\pm$ $16.18$ &78.03 $\pm$ $16.34$&85.34 $\pm$ $1.02$& 83.37 $\pm$ $1.26$ & 71.45 $\pm$ $4.89$& 86.41 $\pm$ $7.75$& 88.39 $\pm$ $6.35$ \\
AUC-ROC weight  & 75.82 $\pm$ $15.82$ &78.46 $\pm$ $16.31$&85.63 $\pm$ $1.54$& 83.50  $\pm$ $1.83$ & 73.94 $\pm$ $7.65$& 86.32 $\pm$ $7.51$& 87.79 $\pm$ $6.92$ \\
F1  sample  & 73.61 $\pm$ $15.33$ &77.77 $\pm$ $16.94$& 84.74 $\pm$ $1.86$& 83.00  $\pm$ $1.12$ & 71.80 $\pm$ $4.92$& 86.23 $\pm$ $7.37$ & 87.81 $\pm$ $6.96$ \\
F1 macro   & 73.65 $\pm$ $15.68$ &77.81 $\pm$ $16.44$& 73.73 $\pm$ $2.22$& 72.88 $\pm$ $1.73$ & 70.17 $\pm$ $5.47$& 86.12 $\pm$ $7.34$& 78.03 $\pm$ $2.05$ \\
F1 micro    & 73.68 $\pm$ $15.48$ &77.85 $\pm$ $16.62$ & 84.80 $\pm$ $1.94$& 82.96 $\pm$ $1.07$ & 71.82 $\pm$ $5.04$& 85.12 $\pm$ $7.20$ & 87.82 $\pm$ $6.90$ \\
F1 weight   & 73.66 $\pm$ $15.43$ &77.82 $\pm$ $16.66$ &84.67  $\pm$ $1.95$& 82.01 $\pm$ $1.44$ & 72.46 $\pm$ $5.08$& 86.57 $\pm$ $7.84$& 85.96 $\pm$ $9.53$ \\
\hline
\multicolumn{8}{|c|}{Binary classification performance of Attention DL networks in $> 100$ epochs} \\
\hline
 Metric  (\%)  & ViT & RGB & RGB-Res & RGB-Den & MHL & MHL-Res & MHL-Den \\
\hline
Recall   & 78.03 $\pm$ $9.01$ & 83.03 $\pm$ $9.97$ & 89.9 $\pm$ $0.84$& 93.73  $\pm$ $1.16$ &72.66 $\pm$ $3.40$&94.89 $\pm$ $3.24$& 95.79 $\pm$ $3.72$\\
Precision    & 81.30 $\pm$ $9.17$ &83.33 $\pm$ $8.85$& 89.95 $\pm$ $0.89$& 93.77  $\pm$ $1.17$ &72.66 $\pm$ $3.43$& 94.90 $\pm$ $3.22$& 95.80 $\pm$ $3.71$\\
AUC-ROC macro  & 81.33 $\pm$ $9.51$ &83.54 $\pm$ $8.90$ & 89.79 $\pm$ $0.84$& 94.46 $\pm$ $1.46$ &75.73 $\pm$ $4.98$& 95.02 $\pm$ $3.73$& 95.97 $\pm$ $3.03$\\
AUC-ROC micro   & 81.23 $\pm$ $9.42$ &83.75 $\pm$ $9.01$ & 89.94 $\pm$ $0.82$ & 94.24 $\pm$ $1.16$ &73.41 $\pm$ $1.69$& 94.96 $\pm$ $3.73$& 95.91 $\pm$ $3.44$\\
AUC-ROC weight  & 81.33 $\pm$ $9.36$ &83.78 $\pm$ $9.15$& 89.92 $\pm$ $0.55$& 94.30 $\pm$ $1.14$ &76.14 $\pm$ $4.83$& 94.95 $\pm$ $3.56$& 95.89 $\pm$ $3.28$\\
F1  sample  & 80.09 $\pm$ $9.20$ &83.81 $\pm$ $9.32$ & 89.95 $\pm$ $0.71$& 93.83 $\pm$ $1.84$ & 72.94 $\pm$ $3.73$& 94.95 $\pm$ $3.32$ & 95.85 $\pm$ $3.36$\\
F1 macro   & 80.03 $\pm$ $9.37$ &83.12 $\pm$ $9.45$ & 89.97 $\pm$ $0.37$& 93.82 $\pm$ $1.65$ & 72.50 $\pm$ $3.18$& 94.57 $\pm$ $3.34$& 95.22 $\pm$ $2.68$\\
F1 micro    & 80.06 $\pm$ $9.19$ &83.21 $\pm$ $9.44$& 90.06 $\pm$ $0.46$& 94.12 $\pm$ $1.47$ & 72.86 $\pm$ $3.65$& 94.67 $\pm$ $3.28$ & 95.82 $\pm$ $3.29$\\
F1 weight   & 80.13 $\pm$ $9.00$ &83.17 $\pm$ $9.90$ & 90.03 $\pm$ $0.41$& 94.15 $\pm$ $1.42$ & 73.08 $\pm$ $3.83$& 95.27 $\pm$ $3.81$& 95.77 $\pm$ $3.36$\\
\hline
\end{tabular}
  \label{tab2}
\end{table*}


Table \ref{st} summarises the results of t-Test analysis. A Bonferroni correction was made due to multiple tests (3 tests for each metric). There was a significant difference in all the metrics between the RGB compared to the DL families (p-value $<0.017$ for significance at the 0.05 level) except recall. There was no significant difference between the RGB and MHL families in all the metrics.

\begin{table}
\caption{t-Test: Two-Sample Assuming Unequal Variances testing for significant differences between networks.}
\centering
\begin{tabular}{ |p{2.75cm}||p{1.cm}|p{1.cm}|p{1.cm}|}
\hline
\multicolumn{4}{|c|}{t-Test analysis of network families } \\
\hline
Network (Metric)  & MHL & RGB & DL  \\
\hline
MHL (Recall) & -  & 0.890  & 0.250 \\
RGB (Recall) & 0.890  & -  & 0.120 \\
DL (Recall) & 0.250 & 0.120  & - \\
\hline
MHL (f1-score) & -  & 0.900   & 0.024 \\
RGB (f1-score) & 0.900  & -  & 0.002 \\
DL (f1-score) & 0.024 & 0.002  & - \\
\hline
MHL (Precision)  & -  & 0.770  & 0.030 \\
RGB (Precision)  &  0.770 & -  &  0.002\\
DL (Precision) & 0.030 &  0.002 & - \\
\hline
MHL (AUC-ROC) & - & 0.900  &  0.007 \\
RGB (AUC-ROC)  &  0.900 & - & 0.007 \\
DL (AUC-ROC) & 0.007 & 0.007  & - \\
\hline
\multicolumn{4}{|c|}{t-Test p-value 0.05} \\
\hline
\end{tabular}
  \label{st}
\end{table}

\subsection{Local and global explainability of the deep learning features}

\begin{figure*} 
\centering
\includegraphics[scale=.645]{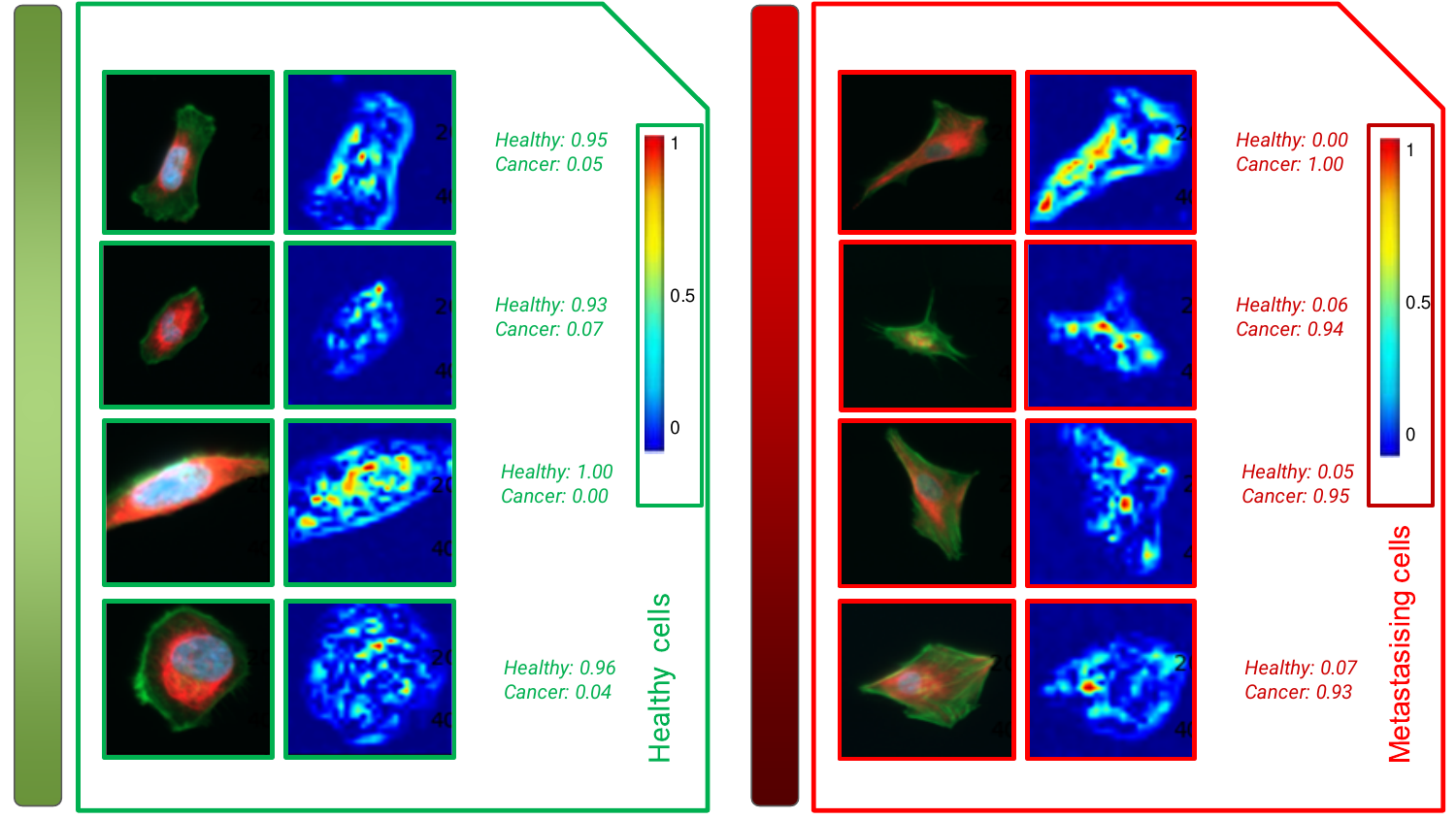}
\caption{Vgg-16 explainable features based on GradCam XAI technique in metastatic and healthy cells. Left column the input healthy (green box) and the metastatic cells (red box). Right Column the GradCam XAI results of the corresponding input image. The colour scale is from 0 to 1 with highest feature contribution the 1 (red). The network's prediction probabilities of each class are given for the corresponding input ( 'Healthy: 0.95' and 'Cancer: 0.05' etc.).}
\label{x1}
\end{figure*}

\begin{figure*}
\centering
\includegraphics[scale=.645]{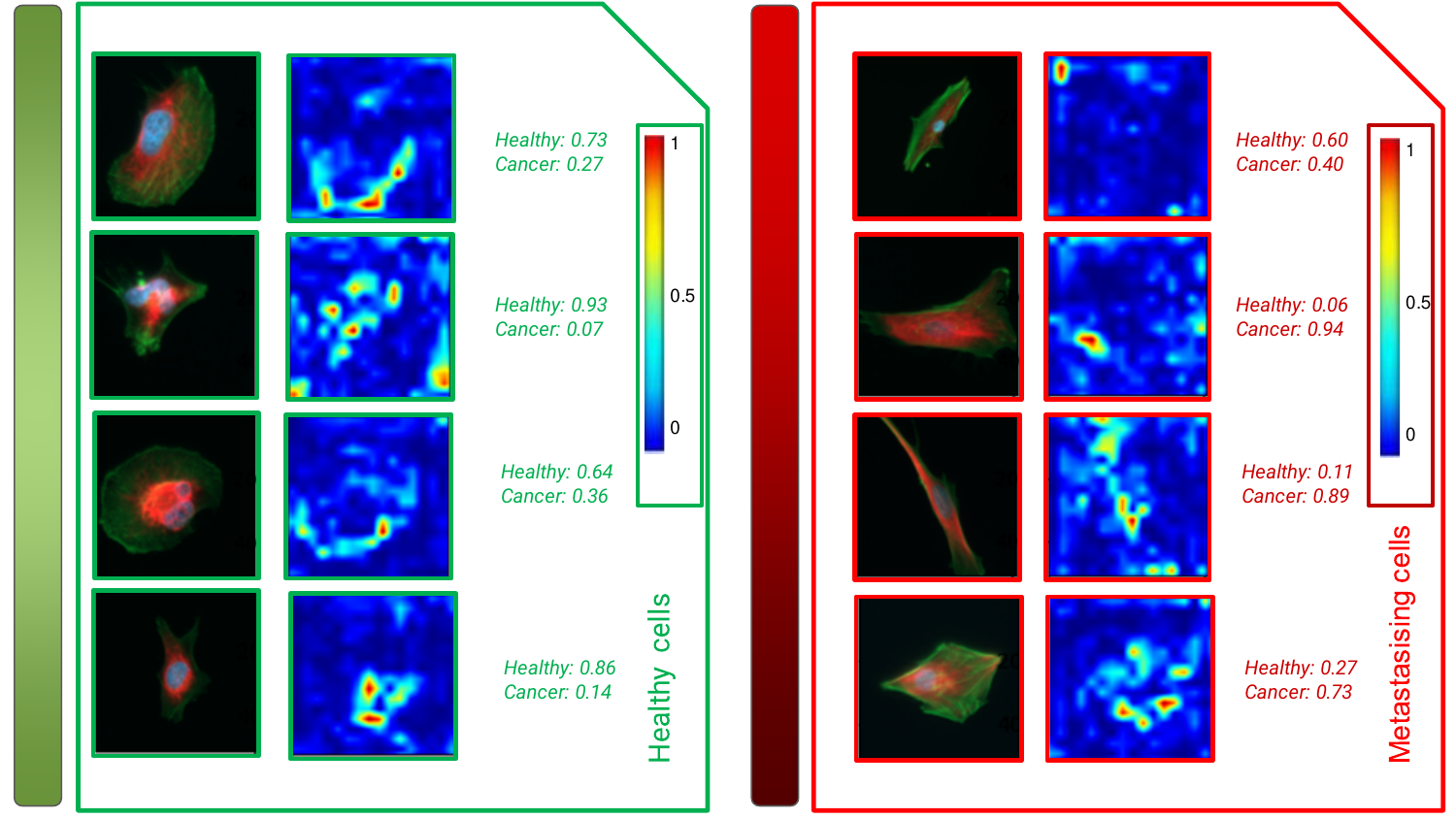}
\caption{DenseNet-121 explainable features based on GradCam XAI technique in metastatic and healthy cells.Left column the input healthy (green box) and the metastatic cells (red box). Right Column the GradCam XAI results of the corresponding input image. The colour scale is from 0 to 1 with highest feature contribution the 1 (red).The network's prediction probabilities of each class are given for the corresponding input ( 'Healthy: 0.73' and 'Cancer: 0.27' etc.).}
\label{x2}
\end{figure*}

\begin{figure*}
\centering
\includegraphics[scale=.645]{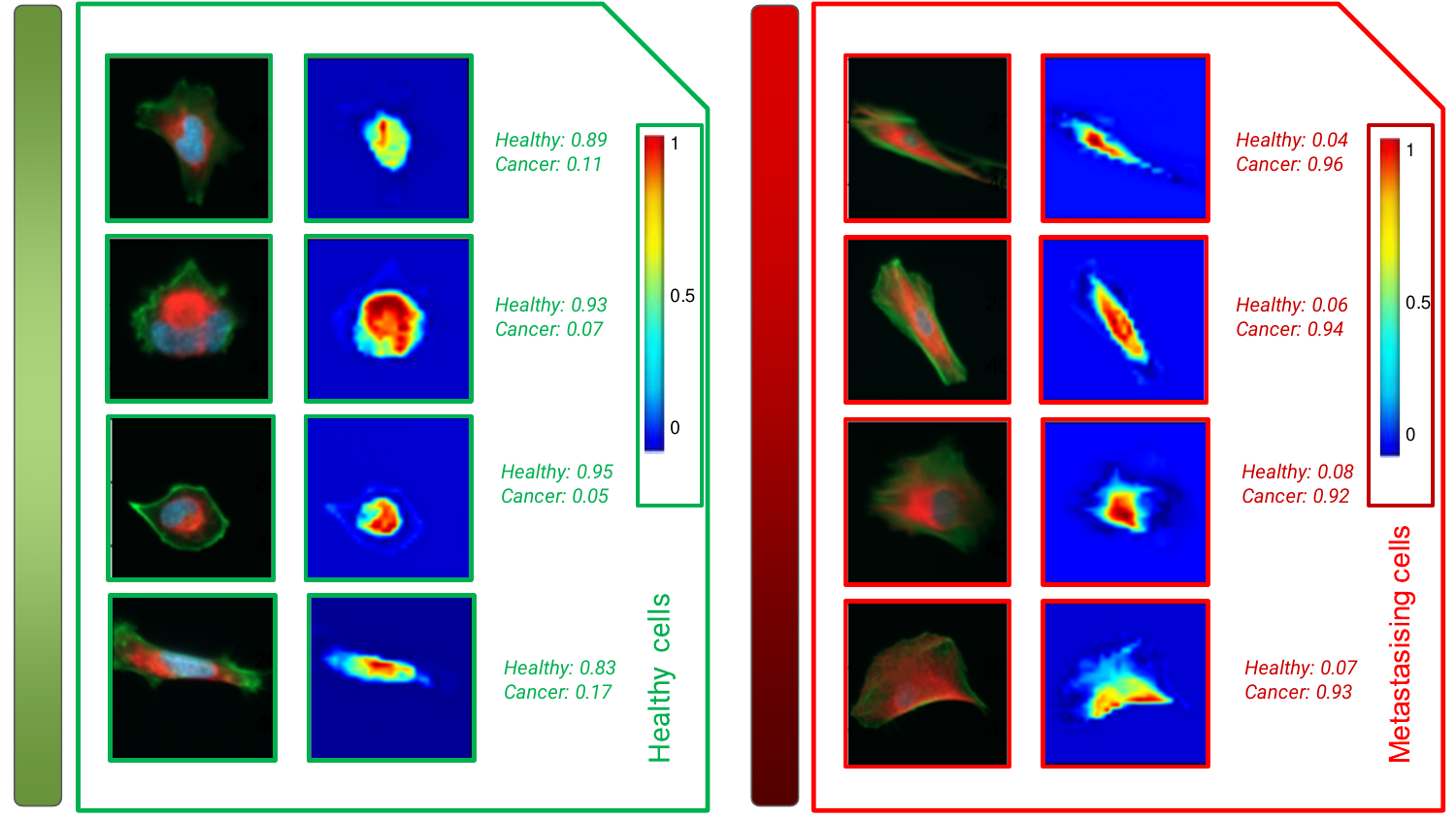}
\caption{ResNet-50 explainable features based on GradCam XAI technique in metastatic and healthy cells.Left column the input healthy (green box) and the metastatic cells (red box). Right Column the GradCam XAI results of the corresponding input image. The colour scale is from 0 to 1 with highest feature contribution the 1 (red).The network's prediction probabilities of each class are given for the corresponding input ( 'Healthy: 0.89' and 'Cancer: 0.11' etc.).}
\label{x3}
\end{figure*}

Fig.~\ref{x1}, \ref{x2}, \ref{x3} show the heatmaps of the last convolutional layer in Vgg-16, DenseNet-121, and ResNet-50 (DL family)  networks. Fig. \ref{x1} shows that the Vgg-16 model is mainly focusing in the vimentin (red channel) of the cells (yellow and red heatmap points) and the total shape of the cells that the F-actin provides (green channel).  
Fig. \ref{x2} shows the GradCam results of DenseNet-121. The network is mainly focusing in the vimentin (red channel) of the cells (yellow and red heatmap points) and the total shape of cells that the F-actin provides (green channel). In some cases the network focuses on some unrelated space points which are out of the cell's region, reducing the prediction accuracy of the model. This can explain the lower performance of DenseNet-121 compared to Vgg-16. Lastly, we present the results of ResNet-121 in Fig. \ref{x3}. ResNet-50 has the best performance compared with the Vgg-16 and the DenseNet-121. Once again the model is strongly focused on the vimentin (red channel) of the cells (yellow and red heatmap points) and the total shape of cells that the F-actin provides (green channel). With very high confidence intervals (probability higher than 0.93) it is focusing on the nuclei region of the cells (blue channel) too.  

\begin{figure*}
\centering
\includegraphics[scale=.645]{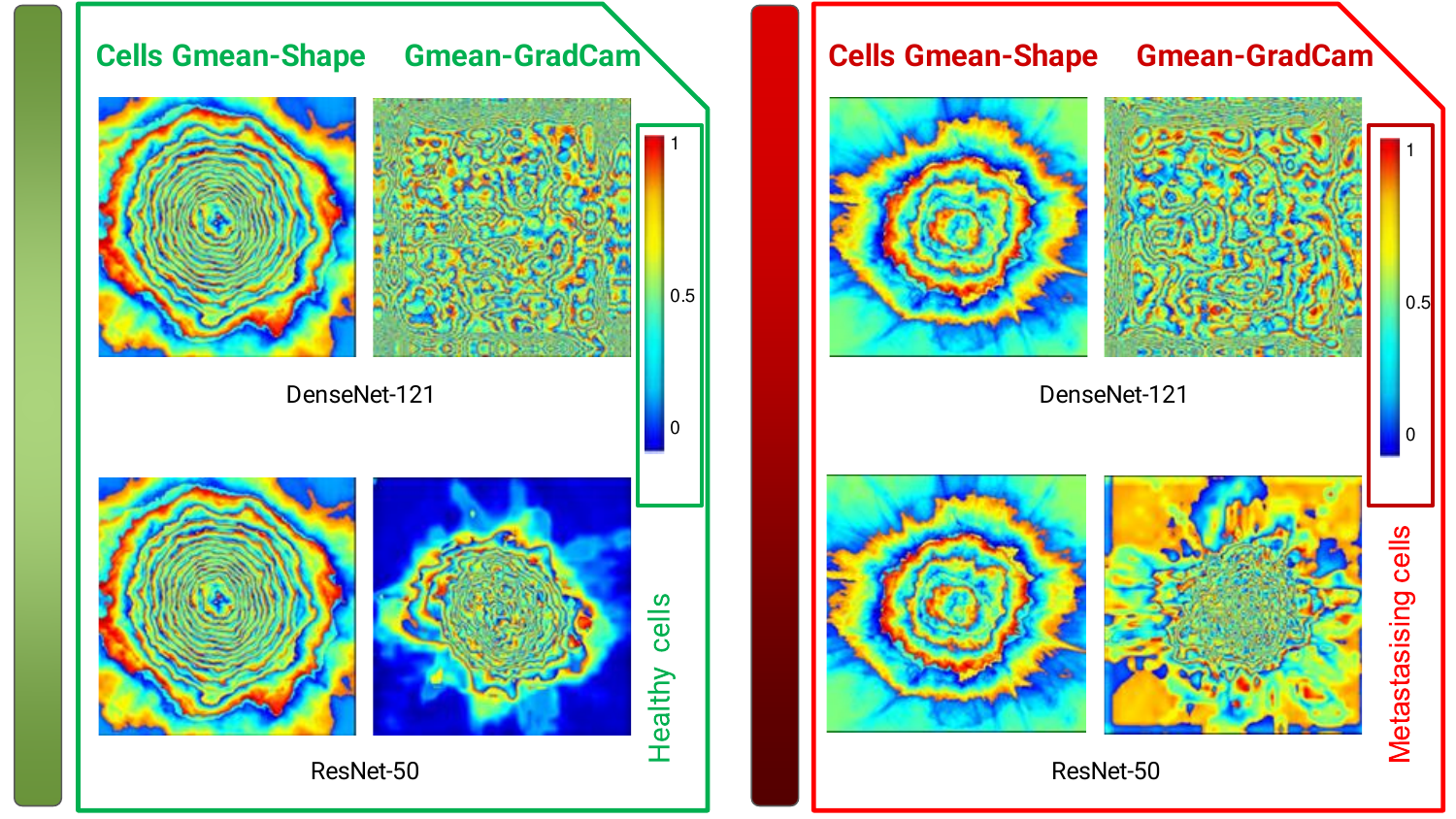}
\caption{Heatmap activation results of weighted geometric mean of GradCam and Shape of the human cells in metastatic and healthy cells. Left column the Gmean-Shape healthy cells input (green box) and the metastatic cells (red box). Right column the Gmean-GradCam XAI results of the corresponding cell's images. The colour scale is from 0 to 1 with highest feature contribution the 1 (red).}
\label{x4}
\end{figure*}

In this study, we delivered a generalised XAI technique that estimates the regions of the images the networks focus attention to, and verifies the validity of the learning process. To deliver this objective we used the weighted geometric mean of all the GradCam results of each cell image (Gmean-Gradcam). Fig. \ref{x4} shows the average shape and Gradcam estimation of all the cells (Gmean-Shape, and Gmean-Gradcam). We present the results of DenseNet-121 and ResNet-50 networks as the worst and best-established networks. We present the weighted geometric mean of the normal and metastasizing cells images, so we can study the differences and variability of the regions of the cell. Fig. \ref{x4} shows the average shape of the cells in normal and metastasizing classes. The healthy cell Gmean-shape is a combination of a large number of curves (more than five high-intensity curves) contrary to the metastasizing where the number of curves is reduced to five. 
\begin{figure*}
\centering
\includegraphics[scale=.665]{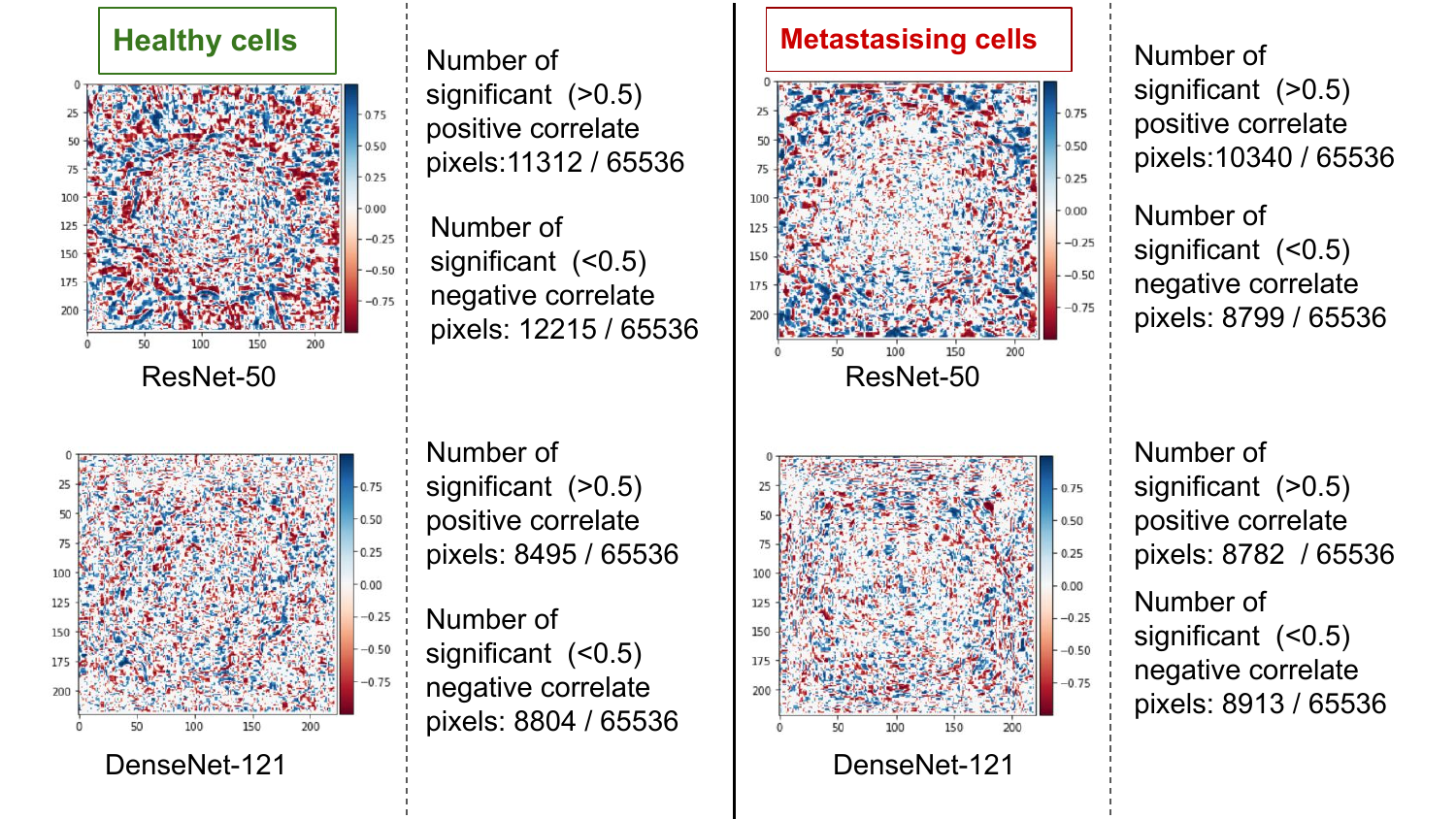}
\caption{Correlation Coefficient of the geometric mean of GradCam and Shape of the human cells in metastatic and healthy cells. Colour scale from -1 (red) to 1 (blue). The significant number was set to higher than 0.5. The positive correlate pixels are those where the pixels significance in both Gmean-Shape and Gmean-GradCam images are in the same orientation (both high or low). The negative correlate pixels are those where the pixels significance in both Gmean-Shape and Gmean-GradCam images are in the invert orientation (one high the other low). }
\label{x5}
\end{figure*}
The number of curves of the healthy cells verify the high variability in the shape of the cells contrary to the low variability of the  metastasizing cells which are more homogeneous. This would be expected, as the metastasizing cells have a reduced capacity to alter their shape based upon cues from the surrounding environment. The Gmean-GradCam shows that ResNet-50 focuses more accurate in the shape and different regions of cells (nuclei, F-actin, vimentin) than the DenseNet-121. This is strengthened by the Fig. \ref{x5}, which shows the correlation coefficient of the normal and metastatic human cells for the Gmean-Shape and Gmean-GradCam of ResNet-50 and DenseNet-121 networks. Resnet-50 has the highest ratio of positive correlate pixels (15.77\%) and the lowest negative correlate pixels (13.42\%) in the metastasizing cells. About the healthy cells ResNet-50 has the highest ratio of positive correlate pixels (17.26\%) and DenseNet-121 has the lowest ratio of negative correlate pixels (13.43\%). The positive correlation ratio includes all the pixels where the significance in both Gmean-Shape and Gmean-GradCam images are in the same orientation (both high or low). The negative correlation ratio includes all the pixels where the significance are in the invert orientation (one high the other low). Therefore, in our study the negative correlation shows that the network did not focusing in the correct shapes of the cells for the classification task, contrary to the positive that shows the correct focusing of the network.
 
\section{Discussion}
In this study, we established and validated a computational large-scale, imaging-based method that can distinguish between normal and metastasizing cells based on the organisation of actin and vimentin filaments in single cells. We further developed multi-attention channels deep learning networks and compared their performance with established networks. Thereafter, we used explainable techniques to analyse the learning patterns the networks used to distinguish between the normal and metastasizing cells, and developed Gmean-GradCam for a more generalised observation about the explainability of the network, based on the significant features that contributed to the final prediction. The Gmean-GradCam showed that the networks focussed accurately on the shape and different regions of the cells (nuclei, F-actin, vimentin). Furthermore, we implemented Gmean-Shape, which showed the variability of the shape of the metastasizing and healthy cells. 

We observed increased homogeneity in the oncogenically transformed, metastasizing cells in comparison to the healthy cells. This is in line with the concept that constitutive activation of oncogenes in different cells activates the same signalling pathways, thereby resulting in similar phenotypes. Cancer cells do not depend on the extracellular factors that regulate the proliferation, motility and shape of healthy cells \citep{m3a}. Therefore, the evidence, found using Gmean-GradCam, that transformed cells have a more homogenous phenotype, compared to normal cells, may be due to transformed and metastasizing cells not adapting to cues in the extracellular environment to which healthy cells respond \citep{dd}.

Our developed multi-head attention channel networks (RGB and MHL families) used the ResNet-50 and DenseNet-121 networks as backbone architectures. The validity of those two networks in the learning process has been verified (GradCam) and thus provides evidence that the multi-attention channel layers of our developed networks increase the performance of the prediction by using global space information of the images \citep{mha}. We therefore claim that the combination of multi-attention and convolutional networks that we used can deliver more robust and accurate results than state-of-the-art deep learning convolutional networks.

We observed a primary focus of the networks on vimentin, and not on actin or the nucleus, which is in line with the current literature and clinical practice, in which the expression of vimentin, but not of actin is used to predict and diagnose invasion and cancer survival. The network showed further a focus on vimentin in the perinuclear cage, the vimentin in the region surrounding the nucleus. Vimentin in this specific location can give mechanical support and protect nuclei from damage during cell migration. This suggests, that vimentin has a role in the mechanical protection of the nucleus that allows cell motility and cancer cell invasion  \citep{sar}. 

The main limitation of this study is the relatively small cohort size. However as the cell protocol and imaging procedure we used are time-consuming, the feasibility of a larger cohort was limited. The main advantage of the imaging is that it provides detailed information of the organisation of actin and vimentin filaments in every single cell. This allows the classification to be successful and the deep learning models to accurately predict the cell type. To capture the variation of the shape of the cells we used augmentation techniques and extend the initial cohort. Our observations further provide evidence of that a multi-head attention channel layer that separately focuses on each (Red, Green, Blue) channel results in more robust and more accurate predictions than methods that focus on the combined channel. In future work, we will collect more images of the cells and evaluate and verify the models in internal and external protocols. We will further determine the features of the networks based on a variate of XAI methods (like LRP etc.), to verify that the focusing pattern we presented in this study (GradCam) is in line with observations of other XAI techniques. Thereafter, we plan to use LRP explainability techniques to study further the prediction patterns of the multi-head attention layers of the MHL and RGB families. 

\section{Conclusion}

To our knowledge, this is the first study that deliver a deep multi-attention channel network to predict and distinguish between metastasizing and healthy human cells with a study of local and global explainable techniques. The multi-head attention layer we developed increased the performance of the DL families and verified that the combination of multi-attention and backbone convolutional networks (MHL-Res, MHL-Den, RGB-Res, RGB-Den) delivers more robust and accurate results than the established deep learning convolutional networks. The RGB family outperform in robustness and accuracy the DL and MHL families. The methods deliver a detailed understanding of how the spatial distribution of the cytoskeleton is defective in, and can cause metastasizing cell, and thereby it can contribute to the future development of novel diagnostic and therapeutic tools against cancer. In addition, our findings highlight that the spatial distribution of vimentin at the micrometre level in cells may be used for future development of diagnostic tools against metastasis. 

\section*{Declaration of Competing Interest}
The authors declare that they have no known competing financial interests or personal relationships that could have appeared to influence the work reported in this paper.

\section*{Acknowledgment}
The authors acknowledge the use of the facilities of the Research Software Engineering Sheffield (RSE), UK and the JADE2 Tier 2 HPC UK system specification and more specifically to Dr. Twin Karmakharm. The authors express no conflict of interest.

\bibliographystyle{model2-names.bst}
\bibliography{refs.bib}

\end{document}